\documentclass[12pt,a4paper]{article}
\usepackage[]{amsmath,amssymb}
\usepackage{graphics,epsfig}
\usepackage{graphicx, color}
\usepackage{geometry}

\begin{document}

\date{} 
\title{Positivity, entanglement entropy, and minimal surfaces}
\author{H. Casini\footnote{e-mail: casini@cab.cnea.gov.ar} 
 \, and M. Huerta\footnote{e-mail: marina.huerta@cab.cnea.gov.ar} \\
{\sl Centro At\'omico Bariloche,
8400-S.C. de Bariloche, R\'{\i}o Negro, Argentina}}\maketitle

\begin{abstract}
The path integral representation for the Renyi entanglement entropies of integer index n implies these information 
measures define operator correlation functions in QFT. We analyze whether the limit $n\rightarrow 1$, 
corresponding to the entanglement entropy, can also be represented in terms of a path integral with insertions on the region's 
boundary, at first order in $n-1$. This conjecture has been used in the literature in several occasions, and specially in 
an attempt to prove the Ryu-Takayanagi holographic entanglement entropy formula.   
 We show it leads to conditional positivity of the entropy correlation matrices, which is equivalent to an infinite 
series of polynomial inequalities for the entropies in QFT or the areas of minimal surfaces representing the entanglement entropy in the AdS-CFT context. 
We check these inequalities in several examples.  No counterexample is found in the few known exact results for the entanglement entropy in QFT. 
The inequalities are also remarkable satisfied for several classes of minimal surfaces but we find counterexamples corresponding to more complicated 
geometries. We develop some analytic tools to test the inequalities, and as a byproduct, we show that positivity for the correlation functions is a 
local property when supplemented with analyticity. We also review general aspects of positivity for large N theories and   
 Wilson loops in AdS-CFT.
\end{abstract}

\section{Introduction}
The positivity of the Hilbert space scalar product gives place to an infinite series of inequalities for the correlators of operators in quantum mechanics. Given a family ${\cal O}_i$, $i=1,...,m$, of operators, a state $\vert 0 \rangle$, and arbitrary complex numbers $\lambda_i$, we have
\begin{equation}
\langle 0 \vert \left(\sum_{i=1}^m \lambda_i {\cal O}_i\right)^\dagger \left(\sum_{j=1}^m \lambda_j {\cal O}_j\right) \vert 0 \rangle=\sum_{i,j=1}^m\lambda_i^*\lambda_j \,\langle 0 \vert   {\cal O}_i^\dagger  {\cal O}_j \vert 0 \rangle \ge 0\,. 
\end{equation} 
Because the $\lambda_i$ are arbitrary, the matrix of correlators has to be positive definite. This is equivalent to
\begin{equation}
\det\left( \{\langle 0 \vert {\cal O}_i^\dagger{\cal O}_j \vert 0 \rangle\}_{i,j=1...m}\right)\ge 0\,,\label{det1}
\end{equation}
for any collection of $m$ operators.   

When the starting point in the description of the theory is given by some numerical functions, these inequalities become specially relevant. In fact, this property allows us to recover the Hilbert space from vacuum expectation values, and plays a central role in Wightman reconstruction theorem in quantum field theory (QFT) \cite{wightman}. 

A remarkable case of a function unexpectedly satisfying inequalities analogous to (\ref{det1}) is given by the traces of integer powers of the local density matrix $\textrm{tr}(\rho_V^n)$, $n=2,3,...$.  Here $\rho_V$ is the vacuum state density matrix $\vert 0 \rangle \langle 0 \vert$ reduced to a region $V$ of the space by tracing over the degrees of freedom outside $V$. 
It is quite surprising these statistical measures give place to operator correlators. The operator behind the identification with correlation functions for $\textrm{tr} \rho^n_{V_i V_j}$ is not $(\rho_{V_i V_j})^{n-1}$, which nevertheless produces the correct expectation values $\langle 0 \left| (\rho_{V_i V_j})^{n-1}\right|0\rangle=\textrm{tr}\rho_{V_i V_j}^n $. Here $(\rho_{V_i V_j})^{n-1}$ is not an operator in the usual sense since it depends on the state. More importantly, inequalities such as (\ref{det1}) further require that $\textrm{tr} \rho^n_{V_i V_j}=\langle {\cal O}_{V_i}^{(n)} {\cal O}^{(n)}_{V_j}\rangle$, for operators ${\cal O}^{(n)}_{V_i}$ and ${\cal O}^{(n)}_{V_j}$ localized in $V_i$ and $V_j$, in some theory, not necessarily the original one in which $\rho$ is calculated. This is not the case of $\rho_{V_i V_j}^{n-1}\neq \rho_{V_i}^{n-1}\rho_{V_j}^{n-1}$.

An operator expectation value interpretation of $\textrm{tr}\rho_V^n$ follows considering the $n$-replicated model. This construction is valid in any quantum mechanical system. Let us consider a system with global state $\left|0\right>$, Hilbert space ${\cal H}={\cal H}_V\otimes{\cal H}_{-V}$, and let $\rho_V=\textrm{tr}_{{\cal H}_{-V}}\left|0\right>\left< 0\right|$. In the replicated model we take $n$ identical copies of this system. The global state is $\left|0^{(n)}\right>=\otimes_{i=1}^n \left| 0_i\right>$, and let a basis of the global Hilbert space of the $i^{th}$ copy be $\{\left|e_a^i\right>\otimes \left|f_b^i\right>\}$, where   $\{\left|e^i_a\right>\}$ is a basis for the Hilbert space ${\cal H}_V^i$ and $\{ \left|f_b^i\right>\}$ is a basis for the one corresponding to the complementary region ${\cal H}_{-V}^i$. Then, defining the unitary operator ${\cal O}^{(n)}_V$ acting on ${\cal H}_V^{(n)}$
\begin{equation} 
{\cal O}^{(n)}_V=\left(\otimes_{i=1}^n\sum_a\left|e^{i+1}_a\right>\left<e^i_a \right|\right)\otimes 1_{{\cal H}_{-V}^{(n)}} \label{chosen}
\end{equation}
it follows that
\begin{equation}
\left<0^{(n)}\right|{\cal O}^{(n)}_V\left|0^{(n)}\right>=\textrm{tr} \rho^n_V\,.
\end{equation}
We also have the splitting relation ${\cal O}^{(n)}_{V_1\cup V_2}={\cal O}^{(n)}_{V_1} {\cal O}^{(n)}_{V_2}$ for non-intersecting $V_1$ and $V_2$. A lattice version of ${\cal O}^{(n)}_V$ has been used in \cite{melko}, where these are called swap operators.  Notice ${\cal O}^{(n)}_{V}$ does not depend on the chosen basis for the Hilbert spaces in (\ref{chosen}).

 The replica trick was first introduced in this context to give an expression of the trace $\textrm{tr} \rho^n_{V}$ in the path integral formulation. This is  
given by the partition function in a $n$-replicated space identified along the $n$ copies of $V$ in such a way one goes from a copy to the other crossing $V$. The 
complete manifold is flat everywhere except at the boundaries of $V$ where it displays a conical singularity of angle $2 \pi n$. For a  detailed description of this 
replica trick see  \cite{replica,replica1,carcal}. These partition functions in non trivial manifolds have been interpreted in the two dimensional case in terms of 
field operator expectation values in \cite{carcal,car}. These are called twisting operators, and are defined to do the job of imposing the correct boundary 
conditions.

It is convenient for later use to express these traces in terms of the ``entanglement'' Renyi entropies $S_n(V)=-(n-1)^{-1} \log(\textrm{tr}\rho_V^n)$, such that 
\begin{equation}
\textrm{tr}(\rho_V^n)=e^{-(n-1)S_n(V)}\,.
\end{equation}
 Allowing for arbitrary real index $n$, the limit $\lim_{n\rightarrow 1} S_n(V)=S(V)=-\textrm{tr}\rho_V \log \rho_V$, gives the entanglement entropy corresponding to $V$.  
 
\begin{figure}
\centering
\leavevmode
\epsfysize=6cm
\epsfbox{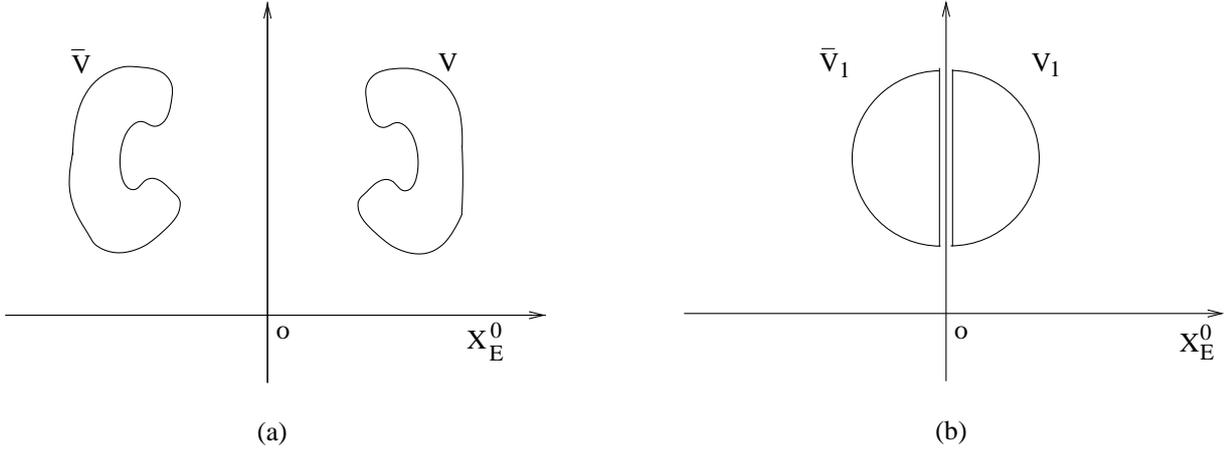}
\bigskip
\caption{(a) Two regions $V$ and $\bar{V}$ in Euclidean space. One is the time-reflected image of the other. (b) If the regions $V_1$ 
and $\bar{V}_1$ are close to the $x^0_E=0$ plane, the common surface near this plane should give an overall numerical factor which drops out from the inequalities. This configuration gives 
information about the set formed by the union $V=V_1\bar{V}_1$ in the limit of no separation. }
\label{parallelas}
\end{figure}

 In the Euclidean version of QFT the positivity inequalities analogous to (\ref{det1}) are called reflection positivity inequalities \cite{rp}. Specifically, for the integer $n$ Renyi entropies these later write
\begin{equation}
 \det \left(\{\, \textrm{tr} \rho^n_{V_i \bar{V}_j}\}_{i,j=1...m}\right)=\det \left(\{e^{-(n-1)\, S_n(V_i \bar{V}_j)}\}_{i,j=1...m}\right) \ge 0\,.\label{det2}
 \end{equation}
Here  $V_i$, $i=1...m$ is a collection of co-dimension one sets included in the half-space of positive Euclidean time, and $\bar{V}_j$ is the Euclidean time-reflected region corresponding to $V_j$. The region $V_i \bar{V}_j$ is just the union of $V_i$ and $\bar{V}_j$ (see figure \ref{parallelas}). A generalized version of these inequalities valid for any quantum mechanical system has been proved in \cite{you} (in a real time formulation of reflection positivity). 

 A short and direct proof of the inequalities (\ref{det2}) follows directly in the usual way reflection positivity is proved with the path integral. 
 This involves splitting 
the path integral for positive and negative time. Writing $\phi^+$ and $\phi^-$ for the restrictions of the field $\phi$ to positive and negative Euclidean time, 
and considering a family of regions $V_i$ in the positive Euclidean time half-space, we have  
\begin{eqnarray}
\sum_{i,j} \lambda_i \lambda_j^* \textrm{tr}(\rho_{V_i \bar{V}_j}^n) ={\cal N}^{-n}\sum_{i,j} \lambda_i \lambda_j^* \int^{V_i \bar{V}_j} {\cal D}\phi\, e^{-S[\phi]}\hspace{7cm}\\
={\cal N}^{-n}\int {\cal D}\phi_0(\vec{x})\,  \left(\sum_i \lambda_i\int^{V_i}_{ \phi^+(0,\vec{x})=\phi_0(\vec{x})} {\cal D}\phi^+ e^{-S[\phi^+]}\right) \left(\sum_j  \lambda_j^* \int^{\bar{V}_j}_{ \phi^-(0,\vec{x})=\phi_0(\vec{x})} {\cal D}\phi^- e^{-S[\phi^-]}\right) \,.\nonumber
\end{eqnarray}  
These are integrals in $n$ replicated space, and the superscript $V$ means the appropriate boundary conditions take place on $V$. The normalization factor ${\cal N}=\int {\cal D}\phi \,e^{-S[\phi]}$ is the one copy path integral without boundary conditions on $V$. The fields $\phi_0$, $\phi$, $\phi^+$ and $\phi^-$ have $n$ independent components. $\phi_0(\vec{x})$ is the common value of $\phi^+(t,\vec{x})$ and $\phi^-(t,\vec{x})$ for $t=0$. Provided the Euclidean action has the time reflection symmetry $S[\phi(t,\vec{x})]=S[\phi(-t,\vec{x})]^*$, a change of field variables $\phi(t,\vec{x})\rightarrow \phi(-t,\vec{x})$ in the path integral shows the two terms in the brackets are conjugate of each other, and the result is positive.    

In QFT the operators ${\cal O}^{(n)}_V$ are localized in $V$, but they do not really depend on the shape of $V$ but only on its boundary $\partial V$.  Different $V$ with the same boundary give place to the same boundary conditions of the path integral. In the real time interpretation this 
is just causality: $\rho_V$ and $\rho_{V^\prime}$ are the same if the spatial sets $V$ and $V^\prime$ have the same boundary because in that case these are equivalent Cauchy surfaces \cite{minkow}. A unitary transformation between the variables on these two surfaces amounts to an irrelevant  change of basis in the Renyi operators (\ref{chosen}). 

The hypersurfaces $V$ have two possible sides. Crossing $V$ from one side, the change of copies in the replicated manifold is in the sequence $1\rightarrow 2\rightarrow 3...$ or in 
the opposite direction $1\leftarrow 2\leftarrow 3...$, according to the chosen orientation for $V$.  If the orientation is determined by a normal vector to the surface $V$, 
in the definition of $\bar{V}$ it is assumed that the orientation vector has changed by a time reflection. We can take advantage of this to obtain inequalities for single 
component sets $V$. Arranging regions $V_1$ close to the surface $t=0$, $V_1\bar{V}_1$ approaches a one component region $V$ (see figure 1b). The sides along the $t=0$ plane have opposite orientation and their contribution  
should decouple from the inequality (\ref{det2}) because the nearby operators should have an OPE proportional to the identity.  

The entropy is obtained from the Renyi entropies in the limit $n=1$. This requires an analytic extension in $n$ from the knowledge of $S_n$ for $n$ integer. This 
extension is unique under reasonable assumptions (see \cite{car, solre}). However, in general there is no path integral representation of $S_n$ for general $n$. The naive 
approach would be  imposing boundary conditions in a manifold, with conical angle $2 \pi n$, for non integer $n$, in the boundary of $V$, and such that for 
integer $n$ one recovers the above mentioned replicated manifolds. Unfortunately this cannot be generally achieved using manifolds which are flat outside $\partial V$. 
A notable exception is the case where there is a ``rotational'' symmetry  with the angle $2 \pi n$ as the parameter of a one dimensional symmetry group 
(see \cite{holoc1} for a nice account of these problems). This is the case of $V$ being the half space in a general QFT, or a sphere in a conformal QFT. 
In curved space-times this symmetry appears typically when the boundary of $V$ is a Killing horizon.      

Another example in which thinking in terms of small deficit angle in the path integral is justified is Solodukhin's formula for the logarithmically divergent term in the entropy for conformal theories and general regions in 
four dimensions \cite{solo}. This is given as a linear combination of two dimensionless,  conformally invariant, and local terms, which are integrals of polynomials 
of the extrinsic curvature on $\partial V$. This general geometric structure of the coefficient of the logarithmic term can be ascribed on general 
grounds to the nature of the ultraviolet divergences in the entropy. The coefficients of these geometrical terms are linear in the two trace anomaly coefficients of the CFT. The trace anomaly coefficients are defined for smooth metrics, hence it is apparent 
the necessity of the limit of small conical defects to establish a connection with the entanglement entropy. For the case where $V$ is a sphere, the coefficient 
turns out to be  proportional to the coefficient of the Euler density in the trace anomaly. This has been later proved to be the case in any dimensions using 
purely QFT arguments, which rely on the existence of a rotational (conformal) symmetry for the sphere \cite{holoc1,sphere} (see also \cite{sphere1}). For a cylinder in flat space the other anomaly coefficient is selected in Solodukhin's formula, though there is no additional rotational symmetry. However, the logarithmic terms are both local and conformally invariant, and in order to compute them a conformal transformation can be used to map the original geometry into another one in curved space, in which there is no extrinsic curvature, and there is a rotational symmetry around the boundary (at least locally)\footnote{We thank Rob Myers and Sergey Solodukhin for clarification of this point.}. The result for the cylinder has also been found by holographic calculations of entanglement entropy for theories with higher derivative gravity actions \cite{myerscilin}, and confirmed in numerical calculations of the logarithmic term for free massless scalars and fermions in 
four dimensions \cite{cyl}.         

\begin{figure}
\centering
\leavevmode
\epsfysize=5cm
\epsfbox{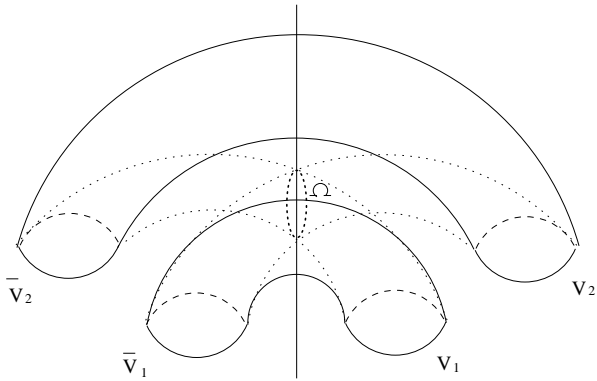}
\bigskip
\caption{Two regions $V_1$ and $V_2$, their time reflected conjugate ones $\bar{V}_1$, $\bar{V}_2$, and the minimal surfaces corresponding to the four combinations $V_i\bar{V}_j$. 
The surfaces corresponding to $V_1\bar{V}_2$ and $V_2\bar{V}_1$ cut at a surface $\Omega$ at $x_E^0=0$, because the points at $x_E^0=0$ are invariant under time reflection. The surface 
$\Omega$ divides these two surfaces in four pieces, and application of the triangle inequality for the minimal areas leads to (\ref{doo}). 
The same argument (though not involving Euclidean time-reflection symmetry) was given in \cite{headrick} to prove strong subadditivity of the holographic entanglement entropy for 
regions lying in a single hyperplane.}
\label{para}
\end{figure}

In more general cases, where there is no path integral representation for non integer $n$, one may still think, as a working hypothesis, that for infinitesimal deficit angles $2 \pi (n-1)$, appropriate for the limit corresponding to the entropy, 
it might be possible to use the path integral formula, and trust it to first order in $(n-1)$. The conical curvature is then placed by insertion of an infinitesimal 
curvature tensor supported on the boundary of $V$. This construction has been used in the literature in several occasions, though up to the present there is no clear 
deductive way to this representation for the entropy in a general region. 

A notable case where the representation of the entanglement entropy in terms of a path integral in a manifold with infinitesimal conical singularities 
is used is Fursaev's  proof of the holographic entanglement entropy \cite{fursaev}, as given by the Ryu-Takayanagi ansatz \cite{ryu}. This ansatz gives 
the entanglement entropy in $V$ in the AdS boundary QFT as $(4 G)^{-1}$ times the area of a minimal surface in the bulk AdS space, with boundary coinciding with $\partial V$,
\begin{equation}
S(V)=(4 G)^{-1} \textrm{min}_{\partial \Sigma=\partial V}{\cal A}(\Sigma)\,.
\end{equation}
We often call the area of the minimal surface directly as ${\cal A}(\partial V)$. This holographic formula for the entropy has passed several tests, 
including giving the correct coefficient of the logarithmically divergent term for the entropy of spheres in any dimension \cite{holoc1,sphere}, the strong subadditive 
inequality for sets in a single spatial hyperplane \cite{headrick}, and the thermal entropy of the boundary in presence of a bulk black hole \cite{nishioka}. 
 However, the missing link between the replicated space for integer $n$ and the small angle 
hypothesis sheds some doubts on one of the assumptions of Fursaev's proof for general geometries, as discussed in \cite{holoc1,hedsolo}.

Thinking in terms of the positivity inequalities, the idea of the small deficit angle path integral representation, with insertions at 
the boundary of $V$ is far from being innocent, and implies a large body of non trivial relations. We have seen above that $e^{-(n-1)S_n(V)}$ 
obeys the correlator inequalities due to its path integral representation. The assumption of the small deficit angle path integral representation  
can be then be rephrased (and generalized) as that $e^{-(n-1)S_n(V)}$ obeys the positivity inequalities at first order in $n-1$ as $n\rightarrow 1$. 
This is equivalent to the matrices $\delta_{ij}-(n-1)S(V_i\bar{V}_j) $ being positive definite for small enough $(n-1)$. As explained in 
section 2 this is the mathematical requirement for the matrices $-S(V_i\bar{V}_j)$ to be conditionally positive definite, or equivalently, 
the matrices of the exponentials $e^{-\lambda S(V_i\bar{V}_j)}$, with any $\lambda>0$,  to be infinitely divisible. Expanding the determinants 
in (\ref{det2}) for small $(n-1)$ we get 
\begin{equation} 
\det(\{S(V_i\bar{V}_{j+1})+S(V_{i+1}\bar{V}_{j})-S(V_i\bar{V}_{j})-S(V_{i+1}\bar{V}_{j+1})\}_{i,j=1,...m-1})\ge 0\,.\label{seiq}
\end{equation}
This is an infinite series of polynomial inequalities in the entropy, one for each polynomial order.  If these conditional positive inequalities where correct the 
entropy itself would have an associated ``almost`` operator, such that $S(V)=\langle {\cal O}(V)\rangle$, $S(V_1 V_2)=\langle {\cal O}(V_1){\cal O}(V_2)\rangle$.
 ${\cal O}(V)$ is an ``almost`` operator, 
in the sense that positivity 
inequalities are not be satisfied by ${\cal O}(V)$ but they are fulfilled by ${\cal O}(V)+c_V$, with $c_V$ a numerical function. This function generally has to be taken to be non translational invariant.
 An example of an almost operator in this sense is given by 
 the free massless scalar field in two dimensions, as will be review later.  

The entanglement entropies have ultraviolet divergences which might in principle obscure the physical content of these inequalities.  Correspondingly, the minimal areas in AdS space diverge due to 
the infinite volume of the surfaces as they approach the asymptotic boundary. 
However, these divergences, being local and extensive along the boundaries $\partial V$, 
cancel out in the generic matrix element in (\ref{seiq}).  
The inequalities can be equivalently written in terms 
of the mutual information $I(U,V)=S(U)+S(V)-S(UV)$, which is finite, as
\begin{equation} 
\det(\{I(V_i,\bar{V}_{j})+I(V_{i+1},\bar{V}_{j+1})-I(V_i,\bar{V}_{j+1})-I(V_{i+1},\bar{V}_{j})\}_{i,j=1,...m-1})\ge 0\,.\label{seiq11}
\end{equation}
The divergences can also be eliminated from (\ref{det2}) by computing the well defined connected  correlators 
\begin{equation}
e^{(n-1) I_n(V_i ,\bar{V}_j)}=e^{(n-1)S_n(V_i)}e^{(n-1)S_n(\bar{V}_j)}e^{-(n-1)S_n(V_i\bar{V}_j)}\,,
\end{equation} 
 with $I_n(U,V)=S_n(U)+S_n(V)-S_n(UV)$. Multiplying the matrix elements in (\ref{det2}) by the factor $e^{(n-1)S_n(V_i)} e^{(n-1)S_n(\bar{V}_j)}$ does not change the positive definite character of the matrix. 

The linear inequality in (\ref{seiq})(with $m=2$) writes
\begin{equation}
2 S(V_1\bar{V}_2) \ge S(V_1\bar{V}_1) +S(V_2\bar{V}_2)\,. \label{doo}
\end{equation}  
It also holds for the integer index Renyi entropies since it coincides with the $m=2$ case in (\ref{det2}). When $V_1 \subseteq V_2$ are on the same hyperplane the inequality (\ref{doo}) coincides with some 
specially symmetric case of the strong subadditive inequality of the entropy. This last inequality has been discussed in the context of minimal areas \cite{hirata1}. 
The reason for the validity of (\ref{doo}) for minimal areas is very simple, and was first discussed in \cite{headrick} in 
relation to the strong subadditive inequality for coplanar regions. It is due to the triangle inequality satisfied by the minimum of the area functional (see figure \ref{para}). 
However, in the present case, the time-symmetric position of the regions in (\ref{doo}) is important for the validity of the inequality, since it 
enforces the relevant minimal surfaces cut each other in order to apply the triangle inequality. Further linear inequalities for the entropy in AdS-CFT have been discussed in \cite{monogamy}.    

We hope that learning whether the $m>2$ non-linear inequalities in (\ref{seiq}) hold will teach us something about the fundamentation of Fursaev's proof and the 
Ryu-Takayanagi proposal for the holographic entanglement entropy. In this paper we take an ``experimental'' attitude and check these inequalities in all (admittedly few) 
known exact calculations of entanglement entropy in QFT. So far, the entanglement entropy can be calculated exactly mostly for free fields in simple geometries 
and these are the cases we have checked. We use either numerical evaluation or analytical tools. Up to what we have checked, we could not find a 
counterexample to these inequalities in the QFT examples. 

Then we check them for minimal areas representing holographic entanglement entropies. We find the 
inequalities are remarkably satisfied in several non trivial cases, though we also find some counterexamples. 
We distinguish two types of counterexamples,  the ones in which the minimal areas in the correlation matrix experiment a phase transition
as a function of the shape of the region, and the ones in which they do not. These phase transitions are related to the fact that the boundary theory is evaluated in a large N and large 
coupling constant limit, and are not expected to occur in the entanglement entropy for ordinary field theories with finite number of degrees of freedom. 
The counterexamples without phase transition violate the ``infinitesimal'' 
inequalities obtained from (\ref{seiq}) for infinitesimally displaced regions.  
The infinitesimal inequalities involve an infinite series of inequalities for the function derivatives of arbitrarily high order. Remarkably, we show the 
positivity relations of correlation functions can be integrated from the infinitesimal inequalities to yield the finite inequalities in regions of analyticity for the correlators. Hence, when the infinitesimal 
inequalities hold in the neighborhood of a fixed region $V$, they automatically hold for all the domain of analyticity. These domains of analyticity are broken 
by phase transitions in the entropy.  Our results give a negative answer for the general validity of conditional positivity 
for the holographic entanglement entropy.      

The theme of minimal surfaces in AdS establishes a natural connection with Wilson loop operators. Maldacena's correspondence gives a geometric 
prescription for the correlators of Wilson loops in a nontrivial interacting theory, the large N limit of ${\cal N}=4$ $SU(N)$ super-Yang-Mills gauge theories 
in four dimensions \cite{wilson}. In Euclidean space and in the the large t'Hooft coupling limit $\lambda \gg 1$ we have 
\begin{equation}
\langle W(C)\rangle =f_0(\lambda,C) e^{-\sqrt{\lambda} \,{\cal A}(C)}\,,
\end{equation}
where $C$ is the (single component) loop, and ${\cal A}(C)$ is the minimal area of a bulk two-dimensional surface whose boundary coincides with $C$. 
This is independent of $N$ at leading order, and $f(\lambda,C)$ is a non exponential correction. 
It is then interesting to look at the way the quantum mechanical positivity property (\ref{det1}) is realized for Wilson 
loops in this context. 

In the next section we review this issue. We start discussing positivity in the broader context of the semi-classical and in large N limits. We recall large N Wilson loops are ``classical operators`` 
in the interpretation of large N theories as classical limits. We show some exponential of classical operators lead to infinite divisible 
correlation matrices. This is analogous to the central limit of probability theory.
 We then discuss large N Wilson loops  
highlighting the deep differences between the entropy and Wilson loop behaviors 
as expansions in $N$ 
and $\lambda$. We show the linear triangle inequality for the minimal areas is enough to satisfy positivity at the leading order in large $N$ and $\lambda$, and 
no higher order polynomial inequalities are in principle required to hold for the minimal areas for Wilson loops.   

In section 3 we prove positivity properties spread on the domain of analyticity of correlation functions. 
In section 4 we analyze known exact examples of entanglement entropy in QFT and check conditional positivity.  
In section 5 we check the geometrical inequalities (\ref{seiq}) for exact solutions of minimal surfaces in AdS. 
Finally, we end with a discussion of the results. 

\section{Positivity in large N and semi-classical limits}

\subsection{Positivity in the semi-classical limit}

The large N limits  can be understood in the same terms as the classical limit of quantum mechanics \cite{yaffe}. Then we first look at 
infinite divisibility in the semi classical limit.

 Consider evaluating the expectation value in ordinary quantum mechanics in the Euclidean framework
\begin{equation}
\langle q(-t_1) q(t_2)\rangle=\frac{\int {\cal D} q(t)\,e^{-\frac{S[q(t)]}{\hbar}}q(-t_1)q(t_2)}{\int {\cal D} q(t)\, e^{-\frac{S[q(t)]}{\hbar}}}\,.
\end{equation}
By reflection positivity this has to be a positive definite kernel for positive $t_1,t_2\ge 0$.  
Let the minimum of the action be attained for some value $q(t)=q_{0}$. 
In the $\hbar\rightarrow 0$ limit the action can be expanded around the classical minimum. Writing the deviations from the classical value as  $x(t)=q(t)-q_0$ we have
\begin{eqnarray}
S&=&S_0+\frac{1}{2}\int dt_1\, dt_2\,  x(t_1)\kappa(t_1,t_2) x(t_2)+{\cal O}(x(t)^3)\,,\\
\kappa(t_1,t_2)&=&\left.\frac{\delta^2 S[q(t)]}{\delta q(t_1)\delta q(t_2)}\right|_{q(t)=q_0}\,.
\end{eqnarray}
 The most relevant part of the functional integral in the limit of small $\hbar$ comes from  values of the fluctuations  $x(t)\sim\sqrt{\hbar}$. Then we have
 \begin{equation}
\langle q(-t_1) q(t_2)\rangle=q_0^2+\hbar\, \kappa^{-1}(-t_1,t_2)+{\cal O}(\hbar^{3/2})\,.\label{coco}
\end{equation}
Positivity for this correlator is trivial to the zero order, but to first order it implies conditional positivity of the kernel
$\kappa^{-1}(-t_1,t_2)$
 \begin{equation}
 \int dt_1\,dt_2\,\alpha(t_1)^* \alpha(t_2)\,\kappa^{-1}(-t_1,t_2)\ge 0\,,\hspace{1.2cm} \int dt\, \alpha(t)=0\,.\label{refiso} 
 \end{equation}
The condition on the possible test functions in (\ref{refiso}) is necessary in order to have an inequality which does not involve the constant zero order term in the correlator (\ref{coco}).  
 
In more generality, a matrix $b_{ij}$ is said to be conditionally positive if
\begin{equation}
x^*_i b_{ij} x_j\ge 0\,,\hspace{2cm}\sum_i x_i=0\,.
\end{equation}
That is, it is positive for any vector $x_i$ satisfying $\sum_i x_i=0$. In the language of QFT $b_{ij}$ is almost a correlator, since it satisfies 
the positivity inequalities except for a one dimensional subspace. Equivalently, $b_{ij}$ is conditionally positive if there is a vector $\xi_i$ such that
\begin{equation}
b_{ij}+\xi_i+\xi_j\label{tercio}
\end{equation}
is positive definite. Another equivalent condition is that the matrix \cite{matrices1}
\begin{equation}
b_{i,j}+b_{i+1, j+1}-b_{i+1, j}-b_{i,j+1}\,\hspace{2cm} i,j=1...m-1
\end{equation}
is positive definite.

Another aspect of positivity in the semi-classical limit is shown using correlation functions of delta function operators $\delta(q(t_i)-q_i)$. These force the paths in the functional integral to pass through the point $(t_i,q_i)$. As a result, in the classical limit, the correlation functions are dominated by exponentials of the classical action
\begin{equation}
\langle \delta(q(t_i)-q_i) \delta(q(t_j)-q_j)\rangle =c\, e^{-S_{\textrm{cl}}((t_i,q_i),(t_j,q_j))/\hbar} (1+{\cal O}(\hbar))\,.
\end{equation} 
Then the reflection positivity inequality for two points boils down to the triangle inequality for the classical action functional, in a way analogous to the one shown in figure (\ref{para}) for the minimal surfaces. We have
\begin{equation}
\det \left(
\begin{array}{cc}
 e^{-S_{\textrm{cl}}((-t_1,q_1),(t_1,q_1))/\hbar} & e^{-S_{\textrm{cl}}((-t_1,q_1),(t_2,q_2))/\hbar} \hspace{3cm}\\
e^{-S_{\textrm{cl}}((-t_2,q_2),(t_1,q_1))/\hbar} & e^{-S_{\textrm{cl}}((-t_2,q_2),(t_2,q_2))/\hbar}
\end{array}\right)\ge 0 \,,
\end{equation}
which is equivalent to
\begin{equation}
S_{\textrm{cl}}((-t_1,q_1),(t_1,q_1))+S_{\textrm{cl}}((-t_2,q_2),(t_2,q_2))\le 2 S_{\textrm{cl}}((-t_1,q_1),(t_2,q_2))\,.\label{24}
\end{equation}
 The triangle inequality is immediate for a real Euclidean action, containing at most first 
derivatives of the fields. 
There is a possible failure of the triangle inequalities for higher derivative theories due to boundary terms in non smooth surfaces which are required by the proof. This is related to 
the failure of these theories to be unitary. 

It is not difficult to convince oneself that the triangle inequality for the euclidean action is all what is required for the positivity of correlator matrices involving several delta function operators, to leading exponential order in $\hbar$. That is, 
in the $\hbar\rightarrow 0$ limit, the inequalities for higher order determinants for delta function operators are satisfied provided the two-dimensional determinant inequality holds, since it is the only one homogeneous in $\hbar$.

\subsection{Infinite divisibility, the central limit and free fields}

In this subsection we comment on the relation between conditional positivity and another property called infinite divisibility, which is related to the central limit theorem. 

Infinite divisible property is inspired by Schur's theorem for positive definite matrices: If $\{a_{ij}\}$ is positive definite (it is hermitian and all its 
eigenvalues are positive) 
then the matrix of the integer powers of the elements $\{(a_{ij})^N\}$ is also positive definite. A positive definite matrix $\{a_{ij}\}=\{c_{ij}^N\}$ which is 
equal to the entry-wise $N$-power of another positive definite matrix $\{c_{ij}\}$ for any $N$ (here $\{c_{ij}\}$ depends on $N$) is called infinite 
divisible \cite{matrices}. In QFT, the infinite divisibility of a correlator, means that its fractional powers also satisfy the positivity conditions a correlator 
ought to satisfy. Then, these fractional powers can be thought as correlators for operators in another theory, non necessarily the original one.

It is not difficult to show that the positive definite matrix $\{a_{ij}\}$, $i,j=1...m$, is infinite divisible if and only if the matrix of the logarithms 
of its entries 
\begin{equation}
b_{ij}=\log(a_{ij})
\end{equation} 
is conditional positive \cite{matrices1}.

An example of correlators which are powers of other correlators is given by a QFT which is the tensor product of identical independent 
sectors named by $1,...,N$. Choosing the 
correlators of product operators
\begin{equation}
\langle 0\vert {\cal O}^{(1)\dagger}_i ... {\cal O}^{(N)\dagger}_i\times {\cal O}^{(1)}_j ... {\cal O}^{(N)}_j\vert 0\rangle =\langle 0\vert 
{\cal O}^{(1)\dagger}_i {\cal O}^{(1)}_j\vert 0\rangle...\langle 0\vert  {\cal O}^{(N)\dagger}_i{\cal O}^{(N)}_j\vert 0\rangle\,.\label{gogo}
\end{equation}

Most correlator matrices in QFT are not infinite divisible. In principle, this property should be valid only for special type of theories and special types of 
operators in these theories. Eq. (\ref{gogo}) suggests that in order to have infinite divisibility one has to consider a theory in the limit of a large number 
of degrees of freedom, and some special kind of operators. We show in the next section this is the case of some exponential operators in large N theories.  

Infinite divisibility is also related, and for the same reason which operates in large N theories, to the central limit of probability theory \cite{prob}. 
Take the probability 
distribution for a variable $y=\frac{1}{\sqrt{N}}(x_1+x_2+...+x_N)$ which is proportional to the sum of $N$ independent variables $x_i$, with the same 
distribution $P(x)$. Then the probability distribution $Q(y)$ for $y$ is the $N$-fold convolution 
\begin{equation}
Q(y)=\int dx_1\,...dx_N\, \delta\left(y-\frac{1}{\sqrt{N}}(x_1+...+x_N)\right) P(x_1)...P(x_N)\,.
\end{equation}
The expectation value for the exponential $G_Q(k)=\langle e^{i k y} \rangle$ (called the characteristic function of the distribution) is
\begin{equation}
G_Q(k)=\int dy\, e^{i k y} Q(y)=\langle e^{i \frac{k}{\sqrt{N}} x}\rangle^N=G_P(k/\sqrt{N})^N\,. \label{seei} 
\end{equation}
Then, the infinite divisible case for the characteristic functions corresponds exactly to probability distributions which arise from an infinite sum of variables. 
Under the condition that the variance of $x$ is finite, the limit of the sum of infinitely many variables can only be the Gaussian distribution, which of course has 
infinite divisible characteristic function. This is the central limit theorem. Different central limit distributions are also possible when the
 variance is non finite. 

Notice that the characteristic functions satisfy  positivity relations which are similar to the ones corresponding to two point correlators in QFT:
\begin{equation}
\lambda_i^* G_Q(k_i-k_j)\lambda_j=\int dx\, Q(x) (\lambda_i e^{-i \frac{k_i}{\sqrt{N}} x})^* (\lambda_j e^{-i \frac{k_j}{\sqrt{N}} x})\ge 0\,.   
\end{equation}
Thus, $G_Q(k_i-k_j)$ are positive definite matrices, and (\ref{seei}) means these write as entry-wise $N$ powers of other positive definite matrices. 

Gaussian distributions are analogous to free fields, and the proof of infinite divisibility for the expectation value of exponential operators is direct. We 
have for a free field $\phi(x)$
\begin{equation}
\langle 0\vert e^{i\int dx\, k(x) \phi(x)}\vert 0\rangle=e^{-\frac{1}{2}\int dx\,dy\,k(x) g(x-y)k(y)}\,,
\end{equation}
where $g(x-y)=\langle 0\vert \phi(x) \phi(y)\vert 0\rangle $ is the two point function, and $k(x)$ is an arbitrary function. 
Then 
\begin{equation}
\langle 0\vert e^{i\int dx\, k(x) \phi(x)}\vert 0\rangle=\langle 0\vert e^{i\int dx\, \frac{k(x)}{\sqrt{N}} \phi(x)}\vert 0\rangle^N=
\langle 0\vert e^{i\int dx\, k(x)\frac{1}{\sqrt{N}} (\phi_1(x)+...+\phi_N(x))}\vert 0\rangle\,.
\end{equation}
The first equation shows the expectation value of an exponential operator as a power of the expectation value of a different exponential operator 
in the same theory, and the second equality shows it as an expectation value in a different theory, which is one of $N$ independent free 
fields $\phi_1(x)$,...,$\phi_N(x)$, with the same two point function. 

Any exponential of a free field gives an operator with infinite divisible correlators.  Another example is the Wilson loop operator $e^{i e \oint dx^\mu A_\mu}$ 
for the free electromagnetic field. Notice that the two point function of the free field $g(x-y)$ has to be only conditionally positive, and this allows 
for operators which are not strictly speaking quantum fields. This is the case of a massless scalar field in two dimensions, whose two point 
function $\langle 0\vert \phi(x) \phi(y)\vert 0\rangle\sim -\log\vert x-y\vert$ cannot be positive for all range of coordinates $x,y$. In section 5 
below we show  $-\log\vert x-y\vert$ is a conditionally positive correlation function. 

 Examples of infinite divisible operators are given by the exponentials $e^{\frac{F[q(t)]}{\sqrt{\hbar}}}$ of classical operators divided by $\sqrt{\hbar}$ is the classical limit. We have 
\begin{equation}
 \frac{F[q(t)]}{\sqrt{\hbar}}=\frac{F[q_0]}{\sqrt{\hbar}} + \int dt_1\, \left.\frac{\delta F[q(t)]}{\delta q(t_1)}\right|_{q(t)=q_0} \frac{x(t_1)}{\sqrt{\hbar}}+{\cal O}\left(\frac{x(t)^2}{\sqrt{\hbar}}\right)\,. 
\end{equation}
 Higher order deviations from the classical value in $F[q(t)]$ or the action are suppressed for small $\hbar$. In this limit, the operator will behave as an exponential of an operator linear in a Gaussian field, and we expect infinite divisibility. We have   
\begin{equation}
\langle e^{\frac{F[q(t)]}{ \sqrt{\hbar}}}\rangle= e^{\frac{F[q_0]}{ \sqrt{\hbar}}} e^{\frac{1}{2 }\int dt_1\, dt_2\, J(t_1)(\kappa^{-1})(t_1,\,t_2)J(t_2)} \left(1+{\cal O}(\sqrt{\hbar})\right)\,,\label{pale} 
\end{equation} 
where $J(t)=\left.\frac{\delta F[q]}{\delta q(t)}\right|_{q\equiv q_0}$. Given a collection of functionals $F_i[q(t)]$ with support on the  positive time $t>0$ region, this later requires the positivity of the matrix (to leading order in $\hbar$)
\begin{equation}
\langle e^{\frac{F_i[q(-t)]^*}{ \sqrt{\hbar}}}e^{\frac{F_j[q(t)]}{ \sqrt{\hbar}}}\rangle=e^{\frac{F_i[q_0]^*}{ \sqrt{\hbar}}}e^{\frac{F_j[q_0]}{ \sqrt{\hbar}}}e^{\frac{1}{2 }\int dt_1\, dt_2\, (J_i^*(-t_1)+J_j(t_1))(\kappa^{-1})(t_1,\,t_2)(J_i^*(-t_1)+J_j(t_1))} \,.\label{ww}
\end{equation}
The positivity of the matrix in (\ref{ww}) is equivalent to the conditional positivity of the kernel $(\kappa^{-1})(t_1,\,t_2)$. We have seen this also follows from the positivity of the two point function to first order in $\hbar$ (\ref{refiso}). 

Using functional integrals in phase space, one can also show in the same fashion the infinite divisibility of operators of the form $e^{\frac{F[q,p]}{\sqrt{\hbar}}}$, provided the functional $F[q,p]$ is well behaved in the $\hbar \rightarrow 0$ limit, and that $F[q,p]$ is such that the functional integral converges. 
 The operators such as $F[q,p]$ which in the limit $\hbar \rightarrow 0$ can be expressed as a function in phase space (technically, they have a well defined limit of expectation values in any coherent state) are called classical operators in \cite{yaffe}. Thus, the infinite divisible operators $e^{\frac{F[q,p]}{\sqrt{\hbar}}}$ are not classical operators. They are half way between classical operators of the form $e^{F[q,p]}$ and coherent operators of the form $e^{\frac{F[q,p]}{\hbar}}$.  The vacuum expectation values of operators of the form $e^{\frac{F[q,p]}{\hbar}}$ 
 produce much greater excitations in the $\hbar\rightarrow 0$ limit, and test terms in the action which are of higher order that the Gaussian term. Thus, in general there is no infinite divisibility for these operators. 
  
In the large N limit of interacting theories \cite{ot} a result on infinite divisibility analogous to the one for free 
fields depends on choosing adequate operators that test only Gaussian fields fluctuations in leading approximation.

\subsection{Large N theories as classical limits}
According to the general scheme of \cite{yaffe} the large N limits can be thought as classical limits. This is done by adequately choosing a set of classical variables 
in the large N theory, and identifying the small parameter $\chi$, which plays a role analogous to $\hbar$ in the classical limit, with some inverse power of $N$. The expectation values 
are then computed with a path integral over the classical variables, with a weight given by $e^{-\frac{S_{cl}}{\chi}}$, $S_{cl}$ being a function of the classical variables, 
regular in the $\chi\rightarrow 0$ limit. 

Then, as in the classical limit discussed previously, positivity implies for example the conditional positivity of the connected correlators of classical operators. Equivalently, it also implies the infinite divisibility of operators which are the exponentials of the classical 
operators divided by $\sqrt{\chi}$. While the computation of the action in terms of the classical variables, or evaluating its minimum, can be very difficult in specific 
problems, we do not need this information here. We only need to identify the classical operators. This has been done already in \cite{yaffe} for a large class of models.   

The general scheme (for more details see \cite{yaffe}) starts by defining a coherent group $G$ and a unitary representation $G_\chi$ for each $\chi$. The group $G$ depends 
on the problem. Choosing a base state $\left| 0\right>_\chi$, the coherent states are defined as the states $\left| u \right>=u \left| 0\right>_\chi$ for any $u\in G_\chi$. 
These form an over-complete basis of the Hilbert space. The importance of the coherent states is that their overlap in the $\chi\rightarrow 0$ limit is exponentially small, 
and they can be used to write down a path integral. The classical operators can be defined as the ones which have a well defined limit of  
\begin{equation}
\lim_{\chi\rightarrow 0}\frac{\langle u\left|A\right|u^\prime\rangle_\chi}{\langle u\left.\right|u^\prime\rangle_\chi}\,.
\end{equation}
The interest to us here is that a generating set of the classical operators is obtained from the Lie algebra of the coherence group. If $\Lambda$ belongs to the Lie algebra 
of $G_\chi$, then $(\chi \Lambda)$ is a classical operator. The classical operators are generated by these ones. From this,  natural candidates to infinite divisible 
operators are of the form $e^{i\sqrt\chi\Lambda}$.

The example of the classical limit can be useful here. The coherence group is given by 
operators of the form
\begin{equation}
u=e^{\frac{i}{\hbar}(a p+ b q + c)}.
\end{equation}  
The operators generated by the Lie algebra elements times $\hbar$ are just the algebra of polynomials in $p$, $q$, which are classical operators.

Then, in order to find out which operators are infinite divisible in each model we only need to know the lie algebra of the coherence group.  

For large N $U(N)$-gauge theories, which we are interested here consider, following \cite{yaffe}, a lattice model with link variables $V^\alpha$ which are $N\times N$ unitary matrices, and $\alpha$ labels a oriented link on the lattice. The conjugate momentum $E^\alpha$ is a hermitian matrix, and the commutation relations write
\begin{eqnarray}
\left[E^\alpha_{ij},V^\beta_{kl}\right]&=&\frac{1}{2N} \delta^{\alpha\beta}\delta_{kj}V^\alpha_{il}\,, \\ 
\left[ E^\alpha_{ij},E^\beta_{kl}\right] &=& \frac{1}{2N} \delta^{\alpha\beta}(\delta_{kj}E^\alpha_{il}-\delta_{il}E^\alpha_{kj})\,.
\end{eqnarray}  
The Kogut-Susskind Hamiltonian is invariant under local $U(N)$ transformations, and is given by
\begin{equation}
H=N^2 \tilde{\textrm{tr}} \left(\lambda \sum_\alpha (E^\alpha)^2-\lambda^{-1} \sum_p (V^{\partial p}+V^{\partial \bar{p}})\right)\,.
\end{equation}
Here the t'Hooft coupling constant $\lambda=g^2 N$ is chosen fixed such that the theory has a meaningful large $N$ limit. The normalized trace is $\tilde{\textrm{tr}}=(N^{-1}) \textrm{tr}$, $p$ and $\bar{p}$ represent a lattice plaquette with the two possible orientations. More generally write $V^{\Gamma}$ for the ordered product of the link variables along the lattice path  $\Gamma$.

In this case the Lie algebra of the coherence group is infinite dimensional, and the elements depend on arbitrary closed loops $\Gamma$ on the lattice. It is generated by the operators 
\begin{equation}
i \, N^2 \,\tilde{\textrm{tr}}\, V^\Gamma \,,\hspace{1cm}  i \, N^2 \,(\tilde{\textrm{tr}}\, E^\alpha V^\Gamma+V^\Gamma E^\alpha)\,.
\end{equation}  
Notice that the Wilson loop operators are precisely of the form $\tilde{\textrm{tr}} V^\Gamma$, which are traces of ordered products of the link variables along $\Gamma$. The small parameter is here $\chi=N^{-2}$ \cite{yaffe}. 
Thus, the Wilson loop is a classical operator. This is the reason it has a $N$ independent expectation value (\ref{wl}) in the 
large $N$ limit. Thus we expect the connected correlators to be conditionally positive in the large $N$ limit. 

\subsection{Positivity for Wilson loops in AdS-CFT}

The AdS-CFT prescription for the correlation function of Wilson loops is given by \footnote{We thank Juan Martin Maldacena
 for clarifications on the subject of this section.}
\begin{equation}
\langle W(C_1)W(C_2)\rangle=\langle W(C_1)\rangle \langle W(C_2)\rangle+ \frac{\lambda^2}{N^2}\, f_2(\lambda,C_1,C_2)e^{-\sqrt{\lambda} \,{\cal A}(C_1 C_2)}+
{\cal O}(\lambda^4/N^4)\,,\label{wl}
\end{equation}
This expansion is organized in powers the string coupling $\lambda/N\sim g_s$ and it is assumed the limit of large $N$ is taken first than the limit 
of large $\lambda$. 
  The ${\cal A}(C_1 C_2)$ is the 
minimal area of a surface connecting $C_1$ and $C_2$, which can be either connected or disconnected depending on the relative position of $C_1$ and $C_2$.

Therefore, as expected from the general properties of the large $N$ limit, the connected correlators vanish to leading 
order in $N$. This is very different to what is expected for the entropies of two separated regions. The first term in the large N expansion of the 
entropy $S(V_1V_2)$ is expected to be of order $N^2$ in $d=4$ \cite{ryu} (that is $\sim (G_5^{-1}) R^3$, with $R$ the AdS radius), independent of $\lambda$ (in the large $\lambda$ limit),  and proportional to 
the global minimal area for all surfaces 
with boundary in $\partial V_1 \cup \partial V_2$, 
independently of its topology. Otherwise the mutual information would be always subleading in $N$ with respect to the entropy,
 and this is not possible since when the 
boundaries of $V_1$ and $V_2$ are near the mutual information must tend to twice the entropy.

Thus, while there is direct competition between surfaces 
of different genus for the leading term in the entanglement entropy, this competition  
occurs at higher order in $N^{-1}$ for correlators of Wilson loops. This is because the contribution of surfaces of higher genus to the Wilson loop have additional powers of the string coupling. 
 
 Wilson loops satisfy a reflection positivity relation analogous to (\ref{det2}), 
\begin{equation}
 \det \left(\{\langle W(C_i)W(\bar{C}_j)\rangle\}_{i,j=1...m}\right) \ge 0\,.\label{det22}
 \end{equation} 
Here  $C_i$, $i=1...m$ is a collection of loops included in the half-space of positive Euclidean time, and $\bar{C}_j$ is the Euclidean time-reflected loop
 corresponding to $C_j$. 
The vector tangent to the loop indicating the circulation direction gets multiplied by $-T$, where $T$ is the time reflection matrix. 

Taking the large $N$ limit of the inequality (\ref{det22}), and using the expression (\ref{wl}) for the correlators, we have the conditional positivity of the 
matrix 
\begin{equation}
\left\{\frac{f_2(\lambda,C_i,\bar{C}_j)}{f_0(\lambda,C_i)f_0(\lambda,\bar{C}_j)} e^{\sqrt{\lambda} \left({\cal A}(C_i)+{\cal A}( \bar{C}_j)-{\cal A}(C_i \bar{C}_j)
\right)}\right\}.\label{uh}
\end{equation}
 In the present case, the stronger 
condition of positivity holds for this matrix. To leading exponential order in $\lambda$ this is equivalent to the positivity of $\left\{ e^{-\sqrt{\lambda}
 {\cal A}(C_i \bar{C}_j)}\right\}$ or to the infinite series of inequalities
\begin{equation}
\det \left( \left\{ e^{-\sqrt{\lambda} {\cal A}(C_i \bar{C}_j)}\right\}\right)_{i,j=1,...,m}\ge 0.\label{nue}
\end{equation}
It is not difficult to see 
 that due to the large $\lambda$ limit 
all these inequalities collapse to the linear inequality for two regions, which is the case $m=2$ in (\ref{nue}), and is the only inequality homogeneous in $\lambda$,  
\begin{equation}
2 {\cal A}(C_1\bar{C}_2) \ge {\cal A}(C_1\bar{C}_1) +{\cal A}(C_2\bar{C}_2)\,. \label{doo1}
\end{equation}  
This is analogous to (\ref{doo}). 
 The reason for the validity of (\ref{doo1}) is again the triangle inequality of the minimal areas, the same reason as in the case of holographic entanglement 
entropy \cite{headrick}. Therefore, at least in the large $\lambda$ limit, the positivity of Wilson loops holds provided they are given by exponentials 
of minimal areas with large coefficients, without further requirements on the nature of the surfaces or the metric. In particular no higher order inequalities
 such as (\ref{seiq}) are involved. 

According to (\ref{det22}) for $m=2$, a general version of (\ref{doo1}), the linear reflection positivity inequality for the logarithm of the 
expectation value of the Wilson loops,
\begin{equation}
\log\langle W(C_i)W(\bar{C}_i)\rangle+\log\langle W(C_j)W(\bar{C}_j)\rangle\ge 2 \log\langle W(C_i)W(\bar{C}_j)\rangle
\end{equation}
 is valid in any gauge theory independently of the large N, 
large $\lambda$ limit. This was discussed for example in \cite{quark}. It implies the behavior of the Wilson loop is bounded between a perimeter and an area law. 
A generalization inspired in strong subadditivity has been discussed in \cite{hirata2} for the case of Wilson loops which live in the same two-dimensional plane in 
the AdS-CFT context.  

Since we are considering the large $N$ and large $\lambda$ limit we can think in rearranging the expansion (\ref{wl}) to allow for the large $\lambda$ limit to be taken first than the large $N$ limit. In this case the correlation function of Wilson loops is directly dominated by the exponential of the minimal surface, disregarding of the topology of the surface. It is interesting to note that in this case we are in presence of a different semi-classical limit, with a different parameter $\hbar\rightarrow \lambda^{-1}$ rather than $\hbar\rightarrow N^{-2}$. In this limit the Wilson loop should be considered as a delta function operator in the space of loops rather than a classical operator (in analogy with the discussion at the end of section 2.1), and the correlators are dominated by the exponential of the classical action on the trajectory joining the different points. The inequality (\ref{doo1}) is then analogous to (\ref{24}), which follows from the triangle inequality for the classical action. This explains why positivity of (\ref{nue}) holds in addition to conditional positivity of (\ref{uh}). 

\section{Positivity and analyticity}
In this section we analyze the interplay between analyticity and positivity properties. We have in mind the inequalities for the entropies, but the arguments are generally valid for correlators in any QFT. The positivity inequalities in QFT have different formulations. The  inequalities coming directly from positivity of the scalar product in real time (\ref{det1}) always involve the product of operators at coinciding points, and the distributional character of the correlation functions appears in a fundamental way. We are using here the reflection positivity inequalities in Euclidean time (where, on the other hand, the minimal area prescription holds in the AdS-CFT formulation) which does not involve correlators for coinciding points\footnote{There is also a real time formulation of the reflection positivity inequalities which does not involve products of operators at coinciding points. These ``wedge reflection positivity''  inequalities \cite{you,wedge} write in the present context as (\ref{det2}), but the regions $V_i$ are spatial regions included inside the right wedge $x^1\ge \vert x^0\vert$ in Minkowski space, and the reflection operation is the wedge reflection, $x^{0\prime}=-x^0$, $x^{1\prime}=-x^1$, $x^{j\prime}=x^j$, $j=2...d$.}. 

The conditional positivity of the entropies is given by inequalities (\ref{seiq}). 
  According to the general results of section 2, all the matricial inequalities (\ref{seiq}) can be summarized as a single relation expressing the conditional positivity of the entropy (minimal area) kernel:
\begin{equation}
-\int {\cal D}V_1 \, {\cal D}V_2\,\,  F[V_1] \,S(V_1 \bar{V}_2)\, F[V_2]^*\ge  0\,,\hspace{1.5cm} \int {\cal D}V \,\,F[V]=0\,,\label{loop}
\end{equation}
for any functional on regions $F[V]$ with support on the positive time half-space. The measure in region space is not relevant in (\ref{loop}) since it can be absorbed in $F$.
 Eq. (\ref{seiq}) follows from this one by considering $F[V]$ as a sum of localized delta functionals in the different $V_i$. Eq. (\ref{loop}) follows from (\ref{seiq}) by taking a discrete approximation to $F[V]$ as a sum of delta functions.

In the following, we consider a set of regions described by a finite number of  parameters which live in an open set $D\in R^q$. Let us start with the single parameter case $q=1$. Eq. (\ref{loop}) then writes
\begin{equation}
-\int  dx \,dy\,\,  \phi(x) \,S(x,y)\, \phi(y)^*\ge  0\,,\hspace{1.5cm} \int dx \,\,\phi(x)=0\,,\label{loopa}
\end{equation}   
where $\phi(x)$ is any test function with support in $D$, and we have written $S(x,y)=S(V_1\bar{V}_2)$, for the regions $V_1$ and $V_2$ described by one parameter, $x$ and $y$ respectively (notice $y$ is the parameter corresponding to $V_2$, not to $\bar{V}_2$). In order to fulfill the second condition in (\ref{loopa}) we take $\phi(x)=\partial_x \varphi(x)$. Then, integrating by parts we have 
\begin{equation}
\int  dx \,dy\,\,  \varphi(x) \,f_1(x,y)\, \varphi(y)^*\ge  0\,,\label{loopb}
\end{equation}   
where $f_1(x,y)=-\partial_x\partial_y S(x,y)$. Therefore, the conditional positivity of $S$ becomes the positivity of $f_1$. 

Now, we want to see the implications of (\ref{loopb}) for coordinates in a very small region in the neighborhood of a point $x$. We assume $f_1(x,y)$ is real analytic around a point $(x,x)$. In this neighborhood we write any coordinate as $y=x+\epsilon$. We use again the discrete version of (\ref{loopb}) involving only a finite number $M$ of points, which follows taking $\varphi$ as a sum of localized delta-like wave packets. We have 
\begin{equation}
\sum_{\alpha,\beta=1}^{M} \lambda_\alpha \lambda_\beta^* f_1(x+\epsilon_\alpha,x+\epsilon_\beta)=\sum_{\alpha,\beta=1}^{M} \lambda_\alpha \lambda_\beta^* \sum_{m,n=0}^\infty f_1^{m,n}(x,x) \, \frac{(\epsilon_\alpha)^m}{m!} \frac{(\epsilon_\beta)^n}{n!}\ge 0\,,\label{qpuntoaa}
\end{equation}
for any $M$, $\lambda_\alpha$ and $\epsilon_\alpha$ with $\alpha=1,...,M$. Given a fixed $M$, we choose $\epsilon_\alpha=\epsilon\, \hat{\epsilon}_\alpha$, with all the $\hat{\epsilon}_\alpha$ different. We will later take the limit $\epsilon\rightarrow 0$.
Then, the matrix $(\epsilon_\alpha)^n/n!$ for $\alpha=1...M$, and $n=0...M-1$, is invertible. This follows from the Vandermonde determinant. Hence, for a fixed $\epsilon$ and $\hat{\epsilon}_\alpha$ we can choose the $\lambda_\alpha$ such that  $v_m=\sum_{\alpha=1}^M \lambda_\alpha (\epsilon_\alpha)^m/m!$ for $m=0...M-1$ is any chosen eigenvector of  $\{f_1^{m,n}(x,x)\}_{m,n=0}^{M-1}$, with unit norm. Lets call $\beta$ the corresponding eigenvalue. Since the $\lambda_\alpha$ are at most of order $\epsilon^{-(M-1)}$, the sum $u_m=\sum_{\alpha=1}^M \lambda_\alpha (\epsilon_\alpha)^m/m!$ for any $m\ge M$ is then of order $\epsilon$ at least. 
 We then have 
\begin{eqnarray}
 \sum_{\alpha,\beta=1}^{M} \lambda_\alpha \lambda_\beta^* \sum_{m,n=0}^\infty f_1^{m,n}(x,x) \, \frac{(\epsilon_\alpha)^m}{m!} \frac{(\epsilon_\beta)^n}{n!}
 =\beta+  \sum_{m=0}^{M-1}\sum_{n=M}^\infty f_1^{m,n}(x,x) v_m u_n^*+\nonumber \\
 \sum_{m=M}^{\infty}\sum_{n=0}^{M-1} f_1^{m,n}(x,x) u_m v_n^*+\sum_{m=M}^{\infty}\sum_{n=M}^\infty f_1^{m,n}(x,x) u_m u_n^*  =\beta+{\cal O}(\epsilon)\ge 0\,.\label{arra}
\end{eqnarray}
Taking the limit $\epsilon\rightarrow 0$, this gives $\beta\ge 0$. This is equivalent to say that the matrix $f^{m,n}_1(x,x)$, $m,n=0,...,M-1$ is positive definite. Therefore, we obtain the infinitesimal inequalities
\begin{equation}
\det\left(\{f^{m,n}(x,x)\}_{m,n=0}^{M-1}\right)\ge 0\,\label{deti}
\end{equation}
for all $M$. Hence, the kernel $f_1^{m,n}(x,x)$ is positive definite in the space of all $m,n$.  

Now, for an analytic function, these infinitesimal inequalities imply the ones for finite values of the displacements from the point $x$. Writing (\ref{qpuntoaa}) as
 \begin{equation}
  \sum_{m,n=0}^\infty f_1^{m,n}(x,x) \, \left(\sum_{\alpha}^{M}  \frac{(\epsilon_\alpha )^m}{m!}\lambda_\alpha\right)  \left(\sum_{\beta=1}^{M}\frac{(\epsilon_\beta)^n}{n!} \lambda_\beta\right)^* \ge 0\,,\label{qpunto1}
\end{equation}
we see this inequality holds automatically once the infinitesimal inequalities are valid. 

The equivalence between the infinitesimal and finite inequalities in the convergence neighborhood of the Taylor series have a very important consequence. Once the infinitesimal inequalities hold at a point $x$ they automatically hold at any other point in a neighborhood, and then the infinitesimal inequalities hold in any point inside the convergence radius around $x$. Regions of validity of the infinitesimal inequalities can then be extended in the same way analyticity domains are extended. The infinitesimal inequalities have the same domain as the analyticity domain. The finite inequalities hold (at least) provided the points in the correlator matrix are contained in the convergence radius of the Taylor expansion around some point in this domain. Since correlation functions are usually analytic in QFT, the positivity turns out to be quite a local property.  

From (\ref{deti}) we see the infinitesimal inequalities contain derivatives of the function $f_1$ of all orders. Specifying for the case of a translational invariant entropy which only depends on the Euclidean time variable $x$, $S(x,y)=S(x+y)$, the inequalities write 
\begin{equation}
f_{M}(x)=\det(\{-S^{m+n+2}(x)\}_{m,n=0}^{M-1})\ge 0\,.\label{65}
\end{equation}
 More explicitly the first three inequalities read
 \begin{eqnarray}
f_1(x)&=&-S^{\prime\prime}(x)\ge 0\,,\\
f_2(x)&=&S(x)^{\prime\prime}S(x)^{\prime\prime\prime\prime}-(S(x)^{\prime\prime\prime})^2\ge 0\,, \label{perio}\\
f_3(x)&=&  (S^{(4)}(x))^3+S^{(2)}(x)(S^{(5)}(x))^2
+(S^{(3)}(x))^2 S^{(6)}(x)\label{dietri} \\
&&\hspace{4cm}-S^{(4)}(x)(2 S^{(3)}(x)S^{(5)}(x)+S^{(2)}(x)S^{(6)}(x))
 \ge 0\,.\nonumber
\end{eqnarray}
  Writing $f_0(x) = 1$   
we have a simple recurrence rule for the successive inequalities,
\begin{eqnarray}
 f_{n+1}(x) f_{n-1}(x)&=&f_n^{\prime \prime}f_n(x)-(f_n^\prime(x))^2\,, \label{recurrence}\\
f_n(x)\ge 0\,, \,\,&&\hspace{1.4cm} n=0, 1, 2,... \nonumber
\end{eqnarray}
All these inequalities are clearly independent. It is not difficult to find solutions 
of $f_1(x)>0$ which are not solutions of $f_2(x)>0$ (i.e. $S(x)=\tanh(x)$).

Perhaps it is interesting to see in a simple example how subtle the inequalities (\ref{recurrence}) can turn for $f_n$ with large $n$. Take for example the function $1/x$. Being a correlation function in some QFT, gives place to positive definite correlation matrices. Then taking $f_1(x)=1/x$, $x>0$, all the inequalities are satisfied. Now consider writing $f_1(x)=1/x-a$ for some positive $a$. This is not any more a positive definite function since for large enough $x$ it changes sign. However, for small $x$ the positive definite correlator dominates and one can expect the inequalities are satisfied. A simple calculation shows that $f_n(x)$ changes sign at $x=\frac{1}{n a}$. Thus, for any $x$ there is a negative $f_n(x)$ for large enough $n$. This is in accordance with the above result on the local character of the inequalities for analytic functions: The non positivity of $1/x-a$ can be detected with the infinitesimal inequalities at any $x$, no matter how small. This also shows that the $f_n$ for large $n$ are sensitive to smaller disturbances. On the other hand, if we consider $f_1(x)=1/x+a$ with positive $a$, this is a positive definite function. It then follows that the short distance leading term $f_1(x)\sim 1/x$ has to be a positive definite function by itself. This shows that in checking the inequalities, it is also useful to check positivity for the leading short or long distance approximations.  

When the region depends on more than one variable a similar analysis can be done. Let us write collectively these variables as $x\equiv (x^\mu), \mu=1,...,q$. In a similar way as for the one dimensional case, the conditional positivity of $-S(x,y)$ implies  the positivity of the vectorial kernel given by the derivatives of the entropy, $f_{\sigma \delta}(x,y)=-\partial^x_\sigma \partial^y_\delta S(x,y)$. Expanding this in power series around a point $x$ we get
\begin{equation}
\sum_{\alpha,\beta=1}^{M} \lambda_\alpha^\sigma \lambda_\beta^{\delta *} f_{\sigma\delta}(x+\epsilon_\alpha,x+\epsilon_\beta)=\sum_{\alpha,\beta=1}^{M} \lambda_\alpha^\sigma \lambda_\beta^{\delta *} \sum_{(\mu)(\nu)}^\infty f_{\sigma\delta}^{(\mu),(\nu)}(x,x) \, \frac{(\epsilon_\alpha^1)^{\mu_1}}{\mu_1!}...\frac{(\epsilon_\alpha^q)^{\mu_q}}{\mu_q!} \frac{(\epsilon_\beta^1)^{\nu_1}}{\nu_1!}...\frac{(\epsilon_\beta^q)^{\nu_q}}{\nu_q!}\ge 0\,,\label{qpunto}
\end{equation}
where we have used the multi-index notation $(\mu)=(\mu_1,\mu_2...\mu_q)$, with $\mu_i=0,1,2,..$, for $i=1...q$. 
The strategy is the same as for the one dimensional case. First, we choose the maximum derivative order $p=\textrm{max}(\mu_1+\mu_2+...+\mu_q)$ which will appear in the infinitesimal inequality. This maximum number of derivatives includes a number of different derivative types $(\mu)$ given by the combinatorial $\left(\begin{array}{c} p+q \\ q \end{array}\right)$ \cite{vander2}. Then, we choose the case of $M=\left(\begin{array}{c} p+q \\ q \end{array}\right)$, and the $\lambda_\alpha^\sigma$ such that 
\begin{equation}
v^{\sigma(\mu)}=\sum_{\alpha=1}^{M} \lambda_\alpha^\sigma \frac{(\epsilon_\alpha^1)^{\mu_1}}{\mu_1!}...\frac{(\epsilon_\alpha^q)^{\mu_q}}{\mu_q!}
\end{equation}    
is a normalized eigenvector of $f_{\sigma\delta}^{(\mu),(\nu)}(x,x)$ (in this matrix the derivative types $(\mu)$ and $(\nu)$ belong to the first $M$ types). This can be done  
because it is always possible to choose $\epsilon_\alpha^i$ such that the $M\times M$ multidimensional Vandermonde 
matrix $(\epsilon_\alpha^1)^{\mu_1}...(\epsilon_\alpha^q)^{\mu_q}$ is non singular \cite{vander1}. 
Then, the same reasoning as in (\ref{arra}) leads us to conclude that all the eigenvalues of $f_{\sigma\delta}^{(\mu),(\nu)}(x,x)$ (a $q M\times q M$ matrix) are positive, and this matrix is positive definite. These are the differential inequalities in the multidimensional case. Again, the differential inequalities imply the finite inequalities in the domain of convergence of the Taylor series. This in turn leads to an automatic extension of the validity of the differential inequalities to all the domain of analyticity.  

In order to clarify the discussion, let us be more explicit and look at the example of two variables, $q=2$. For each $p=0,1,2...$ we have that a square matrix of order $q \left(\begin{array}{c} p+q \\ q \end{array}\right)=(p+2)(p+1)$ in the derivatives of $S(x,y)$ is positive definite. However, there may be repeated lines and columns in $f_{\sigma\delta}^{(\mu),(\nu)}(x,x)$ because of the derivatives in $\sigma,\delta$. Since a sub-matrix of a positive definite matrix is also positive definite, we can eliminate these redundant lines. The first case $p=0$ involves  only the type $(\mu)=(0,0)$, and contains, due to the $\sigma,\delta$ derivatives, only first derivatives on each of the entries of $S(x,y)$, 
\begin{equation}-\left(
\begin{array}{cc}
S^{(1,0),(1,0)}(x,x) & S^{(1,0),(0,1)}(x,x) \\
S^{(0,1),(1,0)}(x,x) & S^{(0,1),(0,1)}(x,x)
\end{array}\right)\,. \label{22}
\end{equation}
The positivity of this matrix is a consequence of the linear inequality (\ref{doo}), and always holds for a minimal surface which is the true minimum, not only an extreme of the area functional. 
The second infinitesimal inequality $p=1$ involves the types $(\mu)=(0,0)$, $(\mu)=(1,0)$ and $(\mu)=(0,1)$, and contains one or two derivatives in each entry
\begin{equation}-\left(
\begin{array}{ccccc}
S^{(1,0),(1,0)}(x,x) & S^{(1,0),(0,1)}(x,x) & S^{(1,0),(1,1)}(x,x)&S^{(1,0),(2,0)}(x,x) &S^{(1,0),(0,2)}(x,x) \\
S^{(0,1),(1,0)}(x,x) & S^{(0,1),(0,1)}(x,x) &S^{(0,1),(1,1)}(x,x) & S^{(0,1),(2,0)}(x,x)& S^{(0,1),(0,2)}(x,x)\\
S^{(1,1),(1,0)}(x,x)& S^{(1,1),(0,1)}(x,x)& S^{(1,1),(1,1)}(x,x)& S^{(1,1),(2,0)}(x,x) & S^{(1,1),(0,2)}(x,x) \\
S^{(2,0),(1,0)}(x,x)& S^{(2,0),(0,1)}(x,x)& S^{(2,0),(1,1)}(x,x)& S^{(2,0),(2,0)}(x,x)&S^{(2,0),(0,2)}(x,x) \\
S^{(0,2),(1,0)}(x,x) &S^{(0,2),(0,1)}(x,x)  & S^{(0,2),(1,1)}(x,x) &S^{(0,2),(2,0)}(x,x)  &S^{(0,2),(0,2)}(x,x) 
\end{array}\right)\,.\label{doe}
\end{equation}
We recognize the upper $2\times 2$ sub-matrix is just the matrix (\ref{22}), and the positive definiteness of (\ref{doe}) entails the one of (\ref{22}). However, in the positivity of (\ref{doe}), three more inequalities are added. Hence, it is not enough to check the positivity of the determinant but to check the one of all the eigenvalues is necessary. 

More generally, the analysis of this section can be generalized to the case of infinitesimal inequalities for correlator matrices 
(or conditional positive entropy matrices as we have focused the discussion here), of different operators ${\cal O}_i(x_i)$ evaluated in the neighborhood 
of different points $x_i$ (there may be also different number of parameters determining each $x_i$). The infinitesimal inequalities imply the finite ones inside the 
radius of convergence of the Taylor expansions. This provides a further extension of the inequalities, which propagates inside the multidimensional analyticity domain of 
the correlator matrices. Basically, once positivity holds at some neighborhood of a multi-point $x_i$, the only obstacles for further expansion have to be sourced by 
singularities which break up analyticity.  

\section{Conditional positivity and entanglement entropy in QFT}
In this Section we check conditional positivity of (minus) the entanglement entropy in QFT using numerical and analytical methods. The available exact results for the entropy in QFT are not many, and  
involve mainly free fields in low dimensions and simple geometries \cite{review}. The numerical tests require high precision, typically around $15-20$ good digits for
the $f_5$ defined in the previous Section. This leaves us with a rather restricted scenario, 
both from the analytical and numerical point of view, and precludes the use of approximations such as the ones obtained by putting models in a lattice. 
  
The studied examples are consistent with conditional positivity of $-S(V)$. However, by the above mentioned reasons we are not able to distinguish if this is due to the simple geometries used, or the free character of the models. In this sense, we have to mention two facts. The first one is that the counterexamples found for infinitesimal inequalities on the minimal surfaces in the next Section occur for more complicated geometries for which there are no pure QFT calculations of the entanglement entropy. These involve the entropies of two disjoint circular regions with different radius. The one parameter functions which are our best QFT cases also seem to be conditional positive in the minimal surface case (with the difference that only conformal examples are given by the minimal surfaces, while here we have also massive examples).    

The second commentary is that the case of free fields is up to a certain extent special since in this case the Renyi entropies are also found to satisfy the conditional positivity inequalities while this does not hold for some interacting cases we review below. There is also an indication that the traces $\textrm{tr}\rho_V^n$ with non-integer index $n$ might be positive. If this ``fractionality'' property is correct, extending the positivity to non-integer $n$, it would naturally lead to infinite divisibility of the exponentials of the entropy in free fields. 

However, conditional positivity in the free case does not hold for discrete systems. There is a natural generalization of the reflection positive inequalities applicable to discrete systems \cite{you}. We have applied it for free discrete fermions (though we are not presenting details of this calculation in this paper), and have found counterexamples to the inequalities for some numerical examples using random density matrices in low dimensional Hilbert spaces (though conditional positivity holds for a large fraction of the random examples, specially as the dimensions are increased). Hence, if conditional positivity holds for free QFT, the continuum limit has to be an important ingredient.  

\subsection{One interval in two dimensions: conformal and large mass limits}
The entanglement entropy for an interval of length $L$ in two dimensions and any CFT is given by \cite{replica1,carcal}
\begin{equation}
S(L)=\frac{c}{3}\log(L/\epsilon)\,,
\end{equation}
where $c>0$ is the Virasoro central charge, and $\epsilon$ a short distance cutoff. The logarithm is proportional to the correlation function of massless scalar fields $\langle\phi(L)\phi(0)\rangle\sim -\log(L)$, hence it is a conditionally positive function. We give a direct proof of this in the next Section.

On the opposite extreme to the conformal case, the long distance, large mass leading term for the entropy of an interval in a massive (interacting) theory is also known to have a general expression given by 
\begin{equation}
S(L)=-\frac{1}{8}\sum_i K_{0}(2 m_i L)+\textrm{const}\,,
\end{equation}
where the sum is over the spectrum of massive particles of the theory with masses $m_i$, and $K_0(x)$ is the Bessel function.  This formula has been shown to hold for generic integrable models \cite{car,cari}, and argued to be valid for any massive two dimensional QFT \cite{doyon}. 
 For free fields a direct calculation is in \cite{data,data1}. 
Again, we find this entropy is proportional to a correlator. This is the one of massive free scalar fields in two dimensions. Hence,  it is conditionally positive. In fact, this contribution is positive (rather than conditionally positive) since the massive scalar is a well defined field operator. 

\begin{figure}
\centering
\leavevmode
\epsfysize=6cm
\epsfbox{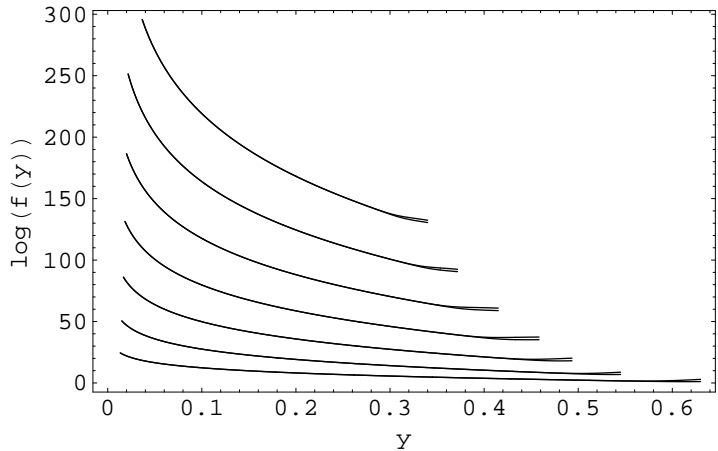}
\bigskip
\caption{ The logarithm of the functions $f_i$, $i=1,...,8$ (from bottom to top) of the parameter $y=m L$ for the fermionic entropy $S(y)$, computed with a series expansion around the origin up to $y^{16}$ and $y^{18}$. The bifurcation between these two approximations at the end of the curves indicates the limit of the region where the series approximation can be trusted. The logarithm is real showing these functions are positive.}
\label{sifi}
\end{figure}

\subsection{ One interval in two dimensions: massive free fields}
Still in $(1+1)$ dimensions, we now study the case of massive scalar (S) and Dirac (D) free fields in a one interval set. The entropy function $S(y)$, with $y=mL$, is known exactly 
and given by \cite{data,data1} (see also \cite{review})
\begin{eqnarray}
S_D(y)&=&\int_0^{\infty}db \frac{\pi}{\sinh(\pi b)^2}S_b^{D}(y)\,,\label{sd}\\
S_S(y)&=&\int_0^{\infty}db \frac{\pi}{\cosh(\pi b)^2}S_b^{S}(y)\,,\label{ssv}
\end{eqnarray}
where
\begin{eqnarray}
S_b^{(D,\,S)}(y)&=&-\int_{y}^{\infty}dt\,\,t\,\,(u_b^{(D,\,S)}(t))^{2} \log(t/y)  \label{sb}\,,\\
u_{(D,\,S)}{^{\prime \prime }}+\frac{1}{y}u_{(D,\,S)}^{\prime} &=&\frac{
u_{(D,\,S)}}{1+u_{(D,\,S)}^{2}}\left( u_{(D,\,S)}^{\prime}\right)
^{2}+u_{(D,\,S)}\left( 1+u_{(D,\,S)}^{2}\right) -\frac{4b^{2}}{y^{2}}\frac{u_{(D,\,S)}}{1+u_{(D,\,S)}^{2}}\,.\label{pds} 
\end{eqnarray}
Equation (\ref{pds}) is a non linear ordinary differential equation of the Painlev\'e type. The difference between the fermionic and scalar cases is exclusively due to the boundary conditions on this equation,
\begin{eqnarray}
u_D(y) &\rightarrow &\frac{2}{\pi} \sinh (b \pi) K_{i 2 b} (y)\,\,\,\,\,\,\,\, \textrm{as} \,\,\,\, y\rightarrow \infty \,,\\
u_{S}(y)&\rightarrow &\frac{2}{\pi} \cosh (b \pi) K_{i 2 b} (y)\,\,\,\,\,\,\,\, \textrm{as} \,\,\,\, y\rightarrow \infty \,.
\end{eqnarray}

The functions $-S_b(y)$ are not conditionally positive, since their large distance approximation $K_{i 2 b} (y)$ does not satisfy the inequalities. However, our numerical tests for the entropies (\ref{sd}) and (\ref{ssv}) using random interval sizes are compatible with conditional positivity. It is not easy to attain excellent numerical precision for these integrals except for the case of the Dirac field where the solution for the differential equation can be obtained by a series expansion around the origin in terms of powers of $y$ and $\log(y)$ \cite{review}. The figure (\ref{sifi}) shows the functions $f_i$, $i=1...,8$, calculated from the fermionic entropy in a region around the origin for the series expansion up to orders $y^{16}$ and $y^{18}$. These functions are positive, what is compatible with the conditional positivity of $-S$. For the scalar case, the expansion for small $y$ starts as $S(y)\sim \frac{1}{3}\log(y)+\frac{1}{2}\log(\log(y))$, which also satisfies the inequalities.

\begin{figure}
\centering
\leavevmode
\epsfysize=6cm
\epsfbox{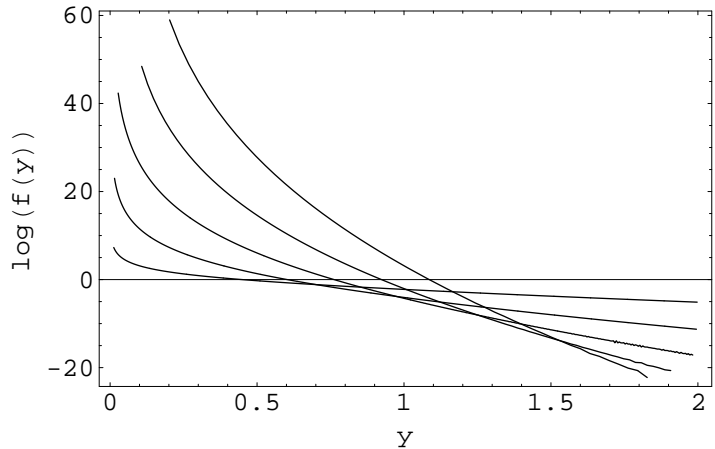}
\bigskip
\caption{ The logarithm of the functions $f_1,f_2, f_3, f_4, f_5$ (from smaller to higher slope) of the parameter $y=m L$ for the fermionic function $S_{ia}$ in the case $a=1/3$. The logarithm is real showing these functions are positive. Similar results hold for different $a$ and for the scalar field.}
\label{painleve-fermion}
\end{figure}

The Renyi entropies are obtained from an analytic extension of $S_b$ for imaginary values of $b$ \cite{review}, 
\begin{eqnarray}
S_n^S(y)&=&-\frac{1}{n-1}\sum_{k=-(n-2)/2}^{(n-2)/2} S_{i\frac{k}{n}}^S(y)\,,\label{rule}\\
S_n^D(y)&=&-\frac{1}{n-1}\sum_{k=-(n-1)/2}^{(n-1)/2} S_{i\frac{k}{n}}^D(y)\,.\label{rulero}
\end{eqnarray}
We checked numerically the differential inequalities (\ref{recurrence}) for the analytically continued functions $S_{i a}$, both for scalar and Dirac fields, and for several values of $a\in (-1/2,1/2)$. In all cases we find these inequalities are satisfied. This points to conditionally positivity of these functions and of the Renyi entropies $-S_n$. The figure (\ref{painleve-fermion}) shows a particular case $a=1/3$. As shown in the figure, all the curves $f_i$ look very similar. This may indicate the conditional positivity in this case could be a consequence of the interplay between the recurrence rules (\ref{recurrence}) for the inequalities and some recurrence relation on  the Painlev\'e differential equation (\ref{pds}).

 The same rules (\ref{sd}), (\ref{ssv}), (\ref{rule}) and (\ref{rulero}) relates the entropy and Renyi entropies for free fields of any region and space-time dimensions, perhaps pointing to the generality of this relation between conditional positivity for the entropies, Renyi entropies, and fractionality in free theories. 

It is interesting to note that the exponentials of the Painlev\'e-related functions $S_{ia}$ and the entropies, would then belong to the class of positive definite infinite divisible functions. Infinite divisible functions which are also probability distributions have been classified in relation to the central limit theorem in probability theory \cite{prob}. 

\subsection{Multiple intervals for a free massless fermion in two dimensions}
Our next example is given by the massless fermion in two dimensions \cite{data}. This is, so far, the only known complete entropy function. The entanglement entropy corresponding to a set $V$ formed by $p$ disjoint intervals $(a_i,b_i)$ on a spatial line, with $a_i<b_i<a_{i+1}$, is
\begin{equation}
S((a_1,b_1)...(a_p,b_p))=\frac{1}{3 }\left( \sum_{i,j} \log | a_i-b_j |-\sum_{i<j}
\log |a_i -a_j |-\sum_{i<j}\log | b_i -b_j |-p\,\log\epsilon \right)\,.\label{fermii}
\end{equation} 
From this entropy function, we find infinite divisibility for the matrices ($e^{- \lambda S(V_i \bar{V}_j)}$) and hence conditional positivity of $- S(V_i \bar{V}_j)$. This follows 
from reflection positivity, since we can write for any $\lambda >0$
\begin{equation}
 \tilde{c}^{p} e^{- \lambda S((a_1,b_1)...(a_p,b_p))}=\langle 0 \vert  :e^{i\sqrt{\frac{2\pi \lambda}{3}} \phi(a_1)}:  :e^{-i\sqrt{\frac{2\pi \lambda}{3}} \phi(b_1)}: ... :e^{i \sqrt{\frac{2\pi \lambda}{3}}\phi(a_p)}: :e^{-i \sqrt{\frac{2\pi \lambda}{3}}\phi(b_p)}:  \vert 0 \rangle \,,\label{fem}
\end{equation}
in terms of vertex (exponential) operators constructed with a free massless scalar field $\phi(x)$ \cite{data}. The right hand side gives the left hand one since we have 
\begin{equation}
\langle 0 \vert e^{i \int dx f(x) \phi(x)}\vert 0 \rangle=e^{\frac{1}{8 \pi}\int dx dy f(x) \log \vert x-y \vert f(y)}\,.\label{quad}
\end{equation} 

 For general free fields the Renyi  
entropies are given in terms of exponentials of operators which are quadratic in the free fields. In this particular case, due to the bosonization (\ref{fem}), they can also be written in terms of exponentials of operators linear in a free scalar field. The infinite divisibility property coincides with the one of these exponential operators as discussed in Section 2 \footnote{The hypothetical theories with extensive mutual information considered in \cite{hypo} also have infinite divisible $e^{-\lambda S}$ given by correlation functions of exponentials of free fields.}. 

In this example, the Renyi entropies $S_n$ are proportional to the entropy, $S_n=\frac{1}{2}(\frac{n+1}{n}) S$, and hence $-S_n$ is conditionally positive. It is interesting to note that in the massive case the Renyi entropies can also be mapped to correlators of exponentials of a scalar field \cite{data}. However, this belongs to the interacting Sine-Gordon theory. Therefore, the infinite divisibility of the two point function discussed in the previous subsection corresponds to the one of correlators of these exponential operators in the Sine Gordon model. 

\subsection{Two intervals in a CFT}
Formula (\ref{fermii}) does not hold for general conformal field theories. Even if no exact result for the entropy is known
 for more than on interval in interacting models there are several interesting exact results for the Renyi entropies of integer $n$ and two intervals for compactified scalars or the Ising model \cite{Calcato, Alba,Furukawa,hea}. Most importantly, the exact results for the Renyi entropies allow us to show that in general the $\textrm{tr}\rho^n$ are not infinite divisible in QFT \cite{you}. An approximation expansion for the entropy in these models are found in \cite{gliozzi}. However, these approximations seem not to be suitable to test positivity. 
 
Conformal invariance implies the entropies of two intervals can be written as
\begin{equation}
 e^{-(n-1)S_n} = k^2 (x(a_2 - b_1 )(b_2 - a_1 ))^{- \frac{c}{6} (n- \frac{1}{n})} F_n (x) ,
\end{equation}
where $F_n (x) = F_n (1 - x)$, $F (0) = 1$, is a function of the cross ratio $x = \frac{(b_1 -a_1 )(b_2-a_2 )}{(a_2 -a_1 )(b_2 -b_1 )}$ , $k$ is a
constant, and $c$ is the Virasoro central charge. 
The functions $F_n (x)$ for the known examples are given by expressions involving
theta functions and their inverses. A simple example is $S_2$ for the critical Ising model which is given in terms of algebraic functions \cite{Alba} 
\begin{equation}
F_2(x)=\frac{1}{\sqrt{2}}\left[\left(\frac{(1+\sqrt{x})(1+\sqrt{1-x})}{2}\right)^{1/2}+x^{1/4}+(x(1-x))^{1/4}+(1-x)^{1/4}\right]^{1/2}\,.
\end{equation} 
We have tested conditional positivity inequalities numerically and find that this Renyi entropy is not conditional positive. However, as expected, the non-linear inequalities (\ref{det2}) hold,
what is consistent with $\textrm{tr}\rho_V^n$ being given by a vacuum expectation value of a twisting operator.
 
We do not know if the infinite divisibility is recovered in the so far unknown
limit $n\rightarrow 1$ for the entanglement entropy. Since the Renyi entropies do not satisfy the inequalities, this case would be a very important test.  
\subsection{Long distance limit for two regions}
In general, for free theories and two sets $A$ and $B$ which are far 
apart we have the mutual information is proportional to the squared of the Dirac or scalar field correlators \cite{review}. Thus, this gives conditional positivity when only the distance $d$ between $A$ and $B$ are changed but not the shapes of the sets. For the case of two intervals in a massive Dirac field in two dimensions we know also the dependence on the shape for large separating distances \cite{Blanco},
\begin{equation}
 I(A, B) \sim\frac{1}{6} m^2 (a_+ b_{-}+ a_{-}b_+ ) K^2_0 (md) +\frac{1}{6} m^2 (a_+ b_+ + a_{-} b_{-} )K^2_1 (md)\,,
\end{equation}
where $a_+$, $b_+$, $a_{-}$ and $b_{-}$ are the projections on the null coordinates of the intervals $A$ and $B$, which are not assumed to lie in a single spatial line but can also be boosted to each other. The positivity of this quantity is proved by noting that the relevant matrix
\begin{equation}
\left(\begin{array}{cc} K_1^2(m d) & K_0^2(m d)\\ K_0^2(m d) & K_1^2(m d)\end{array}\right)
\end{equation}
is the Hadamard square of the correlation function of a massive Dirac field $i \gamma^0 \gamma^1 \langle \Psi(x) \Psi(y)^\dagger\rangle$. This last correlator satisfies the Minkowski space analog of reflection positivity \cite{wedge}. 
\subsection{Angular sectors in 2+1 dimension}
In $(2+1)$ dimensions, we study the infinite divisibility property in the entropy of planar angular sectors of a fixed side length $L$, for free scalar and Dirac fields. We place the angular sectors with common vertex in a plane $x_0, x_1$. We have for the entropy
\begin{equation}
S(\theta)= (2+\theta)\frac{L}{\epsilon}+ s(\theta)\log(\epsilon)+ \textrm{finite}\,. 
\end{equation}
 The only divergent term in the entropy which does not cancel in the conditional positive inequalities is the logarithmic divergent contribution of the vertex angles. This has to satisfy the inequalities independently of the other finite contributions. The problem has then only one parameter.  

\begin{figure}
\centering
\leavevmode
\epsfysize=6cm
\epsfbox{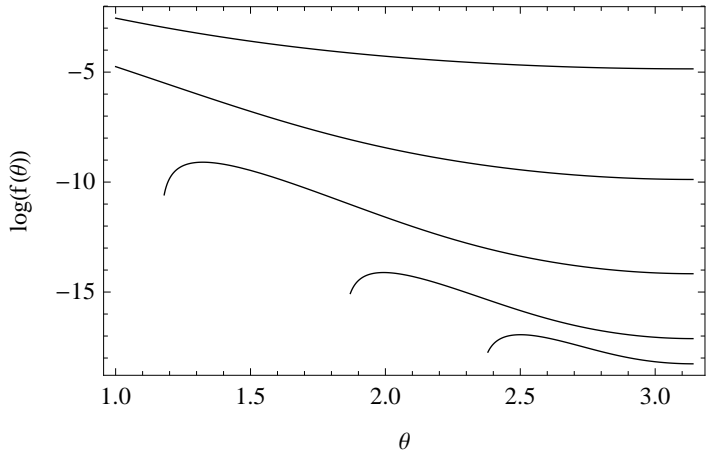}
\bigskip
\caption{The logarithm of the functions $f_1$, $f_2$, $f_3$, $f_4$, $f_5$ (from top to bottom) obtained from the angular entropy function given by the coefficient of the logarithmic term $s(\theta)$ of a plane angular sector for a massless scalar field. We evaluate these functions of the angle $\theta$ as an expansion around $\theta=\pi$ up to order $(\pi-\theta)^{14}$ for $s(\theta)$. These functions are positive in a neighborhood of $\theta=\pi$. For small values of $\theta$ the Taylor expansion for $f_n$ does not give a good approximation, and this is the reason the last three functions go down steeply at some $\theta$. }
\label{fira}
\end{figure}

The checks require to solve with enough precision a coupled set of non linear differential 
equations and integrate the results on one parameter to obtain the logarithmic coefficient of the entropy \cite{angul}.
  Following \cite{angul}, this is done expanding the involved 
functions and differential equations in Taylor series around the value $x=\pi$ up to order $(x-\pi)^{14}$.  
(for details see \cite{angul}). Again, as in the one dimensional case, up to what we have checked, we found the entropy satisfies the conditional positivity inequalities (\ref{recurrence}), as functions of the angle $\theta$ of the angular sector, giving support to infinite divisibility for the exponential of the entropy. The figure (\ref{fira}) shows the functions $f_n$ up to $n=5$ ($6\times 6$ determinants), obtained from $s(\theta)$ corresponding to a massless scalar. For the fermion field we find completely analogous results.

\subsection{Dimensional reduction for free fields}
In higher dimensions, for free fields, some universal terms in the entanglement entropy can be obtained 
via dimensional reduction from results calculated in lower dimensions \cite{review}.
Let us consider sets in $k+d$ spatial dimensions of the form $V=B\times X$, where $B$ is a box on the first $k$ coordinates $x_1,...x_k$, of sides 
lengths $R_i$, $i=1,...,k$, and $X$ is a set in $d$ dimensions. 
The entropy of $V$ in the limit of large $R_i$ is extensive in the sides $R_i$. Thus, we can compactify the directions $x_i$, 
with $i=1,...,k$, by imposing periodic boundary conditions $x_i\equiv x_i+R_i$, without changing the result of the leading extensive term. 
In the limit of large $R_i$, the entanglement entropy $S(V)$ is written in terms of $S(X,m)$, the lower dimensional free entropy function for mass $m$. We have \cite{review}
\begin{equation}
S(V)= \frac{k\,{\cal A}}{2^{k}\pi^{k/2}\Gamma(k/2+1)}\int_0^\infty dp\, p^{k-1}\, S(X,\sqrt{m^2+p^2})\,,\label{ala}
\end{equation}
where the transversal area ${\cal A}=\prod_{i=1}^k R_i$.
The universal terms in $S(V)$ will then come from the ones of $X$ after this integration over the mass. From (\ref{ala}), it is evident that if $S(X,m)$ is 
conditionally positive, this is inherited by $S(V)$ for these special type of regions. For example, the conditional positivity of the entropy of an interval in one spatial dimension is inherited by the entropy of an infinite strip in two dimensions.

\section{Some checks of the geometric inequalities for minimal surfaces}
In this section we check the inequalities which arise from conditional positivity of $- S(V)$, using the holographic prescription (Ryu-Takayanagi ansatz) for the entropy in terms of minimal areas in the bulk AdS space. We find counterexamples for the most complex geometries available, which depend on two parameters. Remarkable, the one parameter functions analyzed are  conditionally positive. 

\subsection{Long strips} 
First consider the case where $V$ is a rectangular strip of width $R$ and length $L$, in the limit of large $L$, in three Euclidean dimensions. Because the result has to be extensive in $L$, and because of conformal invariance, we have
\begin{equation}
S(V)=c_0\frac{L}{\epsilon} -c_1 \,\frac{L}{R}\,,\label{piro}
\end{equation} 
for some constants $c_0>0$, $c_1$, and  a distance cutoff $\epsilon$. 

\begin{figure}
\centering
\leavevmode
\epsfysize=6cm
\epsfbox{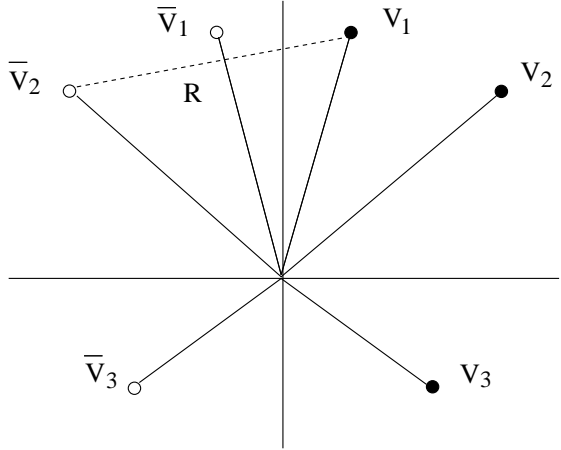}
\bigskip
\caption{Three rectangular regions $V_1$, $V_2$ and $V_3$, and their time reflected images, with common side at the origin. Only the short 
sides of the rectangles are shown. The loop $V_1 \bar{V}_2$ is equivalent to a rectangular loop with side $R$, in the limit of large longitudinal size $L\gg R$.}
\label{fi3}
\end{figure}

In order to obtain the positivity inequalities for single component regions of this kind, we position different regions in the $x^0\ge 0$ space, with the long sides lying 
parallel to the $x^3$ direction. Further, we require one of the vertices to lie at the origin $x=0$ (see figure \ref{fi3}) and all strips to 
have the same length $L$ and orientation. Hence, all regions have a common boundary line. Because of the cancellation of the boundary lines passing on the point $x=0$ belonging to the 
regions and their reflected counterparts, the area is a two point function 
depending on the coordinates $\hat{x}=(x^0,x^1)$ of the free boundary line of each region.
 We now prove the conditional positivity of minus
the function (\ref{piro}), where $R=\vert\hat{x}-\hat{y}\vert$. Notice the conditional positivity condition does not take into account the cutoff 
dependent term in (\ref{piro}). 
 
 We can expect the positivity of $R^{-1}$, and hence the conditional positivity of $-S(V)$, because $R^{-1}$ is a field correlator in two dimensions. 
This requires $c_1\ge 0$ only. The same happens for any power $R^{-\nu}$ with $\nu\ge 0$, since $R^{-\nu}$ are field correlators for any $\nu>0$: these are two point 
correlators of fields with different conformal weight in two dimensional CFT.   
It is instructive to go through the explicit proof in this two dimensional case. We are going to show that the correlators $R^{-\nu}$ are positive for 
all $\nu>0$. This is equivalent to $-\log(R)$ being conditionally positive. This writes 
\begin{equation}
-\int d^2x\, d^2y\, f(x) f(y)^* \log((x^0+y^0)^2+(x^1-y^1)^2) \ge 0\,,\label{tidos}
\end{equation} 
conditioned to $\int d^2x\,f(x)=0$.
 Writing the correlators as a Laplace-Fourier transform
\begin{equation}
\log((x^0+y^0)^2+(x^1-y^1)^2 ))=\int dk_0\, dk_1\, e^{i k_1 (x^1-y^1)-k_0 (x^0+y^0)} g(k_0,k_1)\,,
\end{equation} 
and replacing this in (\ref{tidos}) we have 
\begin{equation}
\int dk_0 \, dk_1\,\vert\hat{f}(k)\vert^2 g(k_0,k_1)\ge 0\,,
\end{equation}
with
\begin{equation}
\hat{f}(k)=\int d^2 x\, e^{-k_0 x^0+i k_1 x^1} f(x)\,,\hspace{1.5cm} \hat{f}(0)=0.
\end{equation}
This last Laplace transform makes sense since the condition for reflection positivity is  $f(x)=0$ for $x^0<0$. Then, all the infinite series of inequalities are 
summarized as $g(k_0,k_1)\ge 0$ for $k\neq 0$. The explicit evaluation of the Laplace anti-transform of the logarithm confirms this expectation
\begin{equation}
g(k_0,k_1)=\frac{\delta(k_0-\vert k_1 \vert)}{\vert k_1\vert}\,.
\end{equation}

For higher dimensions there is a minimal power $\nu$ such that $R^{-\nu}$ is a positive correlation function. This unitarity bound for two point functions in CFT is $\nu\ge (d-2)$. The proof can be done using the Laplace transform as above. Notice that for dimensions $d>2$ this means there is a lowest power for which the two-point correlator can be positive and hence there are no infinite divisible two-point correlation functions. However, this is not the case for extended objects such as surface operators of surface dimension $q$. Then in the conformal case we expect correlators proportional to $e^{L^q/R^q}$, where $L$ is the large size of the branes in the $q$ directions, and $R$ is the separation between two branes. Thus $R$ lives in $d-q$ dimensions. The unitarity bound implies this correlator is infinite divisible ($R^{-q}$ is conditionally positive) only if $q\ge(d-2)/2$. For the case of the entropy $q=d-2$ and we always have infinite divisibility of strips in any dimension. This is also the case of Wilson loops for free electromagnetic field in $d=4$, which has $q=1$.     

\subsection{Angular sectors}
Now consider loops which bound a plane angular sector of angle $\theta$. The area
 as a function of the angle is \cite{angulo}
\begin{equation}
{\cal A}(C)=c_1+c_2 (2+\theta) \frac{L}{\epsilon} - g(\theta) \log(L/\epsilon)\,.\label{pisto}
\end{equation}
 The function $g(\theta)$ is given in parametric form in terms of a variable $x\in (0,\infty)$ as
\begin{eqnarray}
\theta&=& 2 x \sqrt{1+x^2}\int^\infty_0 \frac{dz}{(z^2+x^2) \sqrt{(z^2+x^2+1)(z^2+2 x^2+1)}}\,,\\
g&=& \int^\infty_0 dz\,\left(1-\frac{\sqrt{z^2+x^2+1})}{\sqrt{z^2+2 x^2+1}}\right)\,.\label{jaka}
\end{eqnarray}     
This gives $g(\theta)$ for $\theta$ between $\theta=0$ ($x \rightarrow \infty$) and $\theta=\pi$ ($x\rightarrow 0$). We can extend it to $\theta\in [0,2 \pi]$ by using $g(2\pi-\theta)=g(\theta)$. 

We consider angular sectors on the plane $(x^0,x^1)$, of the same size $L$ with common vertex at $x=0$, and common side along the $x^{1}$ axes. Then, the first two terms in (\ref{pisto}) are not relevant for the inequalities.  These write 
\begin{equation}
\int_{0}^{\pi} d\theta_1\,\int_{0}^{\pi}\,d\theta_2\, u(\theta_1) g(\theta_1+\theta_2) u(\theta_2)^* \ge 0\,,
\end{equation}
for arbitrary $u(\theta)$ with support in $\theta\in [0,\pi]$, and $\int d\theta\, u(\theta)=0$.

\begin{figure}
\centering
\leavevmode
\epsfysize=6cm
\epsfbox{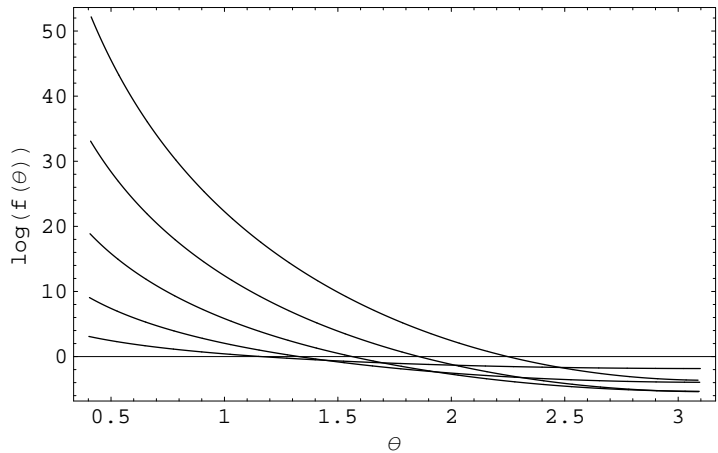}
\bigskip
\caption{The logarithm of the functions $f_i(\theta)$, with $i=1$ to $i=5$ (the curves with higher slope correspond to larger $i$).  These functions are all positive (the logarithms are real). They are symmetric around $\theta=\pi$, $f_i(\theta)=f_i(2 \pi-\theta)$.}
\label{angle}
\end{figure}

If we could write $g(\theta)$ as a Laplace transform (perhaps for a distributional or discrete function $\hat{g}(p)$), 
\begin{equation}
g(\theta)=\int dp\, e^{-p \theta} \hat{g}(p)\,,\label{exu}
\end{equation} 
 we could simplify the inequalities as $\hat{g}(p)\ge 0$ using the same trick as in the previous section. Unfortunately at present we cannot understand whether an expression like (\ref{exu}) is valid. 
 Thus, we can not offer a complete analytical check of these inequalities here.

 However, we have checked numerically the functions $f_n(\theta)$ of the derivatives of $g(\theta)$ which are predicted to be positive by the inequalities (\ref{recurrence}).  Up to what we have checked, these functions are indeed positive (see figure \ref{angle}). We have also checked the positivity of the determinants using up to $6\time 6$ random angles entropy matrices. 

\subsection{Coplanar circles}

We consider two coplanar circles of radius $R_1$ and $R_2$, at a distance $h=h_1+h_2$, where $h_1$ and $h_2$ are the distances of the circles to the plane $x^0=0$. This 
case contains objects determined by two parameters, $(h,R)$. The area for two parallel coaxial circles (rather than coplanar) was calculated in \cite{za} and \cite{za1}. This case can be transformed conformally to the one of the two coplanar circles \cite{tierri}.  Taking out a divergent piece proportional to the 
circles perimeter, the relevant part of the area is conformally invariant.  It is then given as a function of the cross ratio $z=\frac{|x_1-x_4||x_2-x_3|}{|x_1-x_2||x_3-x_4|}$ of four points. We choose $x_1$ and $x_4$ to be the points in the two circles which are at a greatest distance to each other, and $x_2$ and $x_3$ the ones at the minimal distance. Then the cross ratio is
$z = h \frac{(2 R_1 + 2 R_2 + h)}{4 R_1 R_2}\in (0,\infty)$. We have a parametric expression for the area \cite{za1} 
\begin{eqnarray}
{\cal A}&=&-4 \pi \frac{\alpha}{\sqrt{\alpha-1}}\int_0^{\pi/2} \frac{dt}{ (1+\alpha \sin^2(t)+\sqrt{1+\alpha \sin^2(t)})}=-4 \pi \frac{E(-\alpha)-K(-\alpha)}{\sqrt{\alpha-1}}\,,\\
\alpha &=&\frac{1+2 x^2+\sqrt{1+4 x^2}}{2 x^2}\,,
\end{eqnarray}
where $E$ and $K$ are the complete elliptic integrals of the first kind. The parameter $x$ is related to the geometry of the configuration by the following equation involving the cross ratio $z$
\begin{eqnarray}
\frac{1}{2}\log\left(1+2 z+2 \sqrt{  z^2 +z}\right)=x\int_0^{\arccos\left(\frac{\sqrt{4 x^2+1}-1}{2 x}\right)}d\phi\, \frac{\sin^2(\phi)}{\sqrt{\cos^2(\phi)-x^2 \sin^4(\phi)}} \label{largui}\\
\hspace{1cm}= \frac{\sqrt{2}x\left(K\left(-\frac{1+2 x^2+\sqrt{1+4 x^2}}{2 x^2}\right)-\Pi\left(\left.-\frac{1+\sqrt{1+4 x^2}}{2 x^2}\right|-\frac{1+2 x^2+\sqrt{1+4 x^2}}{2 x^2}\right)\right)}{\sqrt{\sqrt{1+4 x^2}-1}}\nonumber\,,
\end{eqnarray} 
where $\Pi(x\vert y)$ is the complete elliptic integral of the third kind.

\begin{figure}
\centering
\leavevmode
\epsfysize=6cm
\epsfbox{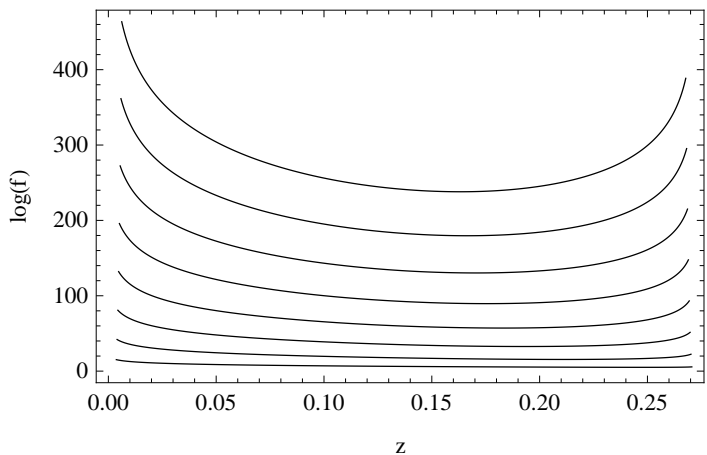}
\bigskip
\caption{The logarithm of the functions $f_i(z)$ for $i=1...8$ (from bottom to top) corresponding to coplanar circles of equal radius (here $R_1=R_2=1$). They are all positive in the range $z\in (0,z_0)$, corresponding to connected minimal surfaces.}
\label{paraq}
\end{figure}

\begin{figure}
\centering
\leavevmode
\epsfysize=6cm
\epsfbox{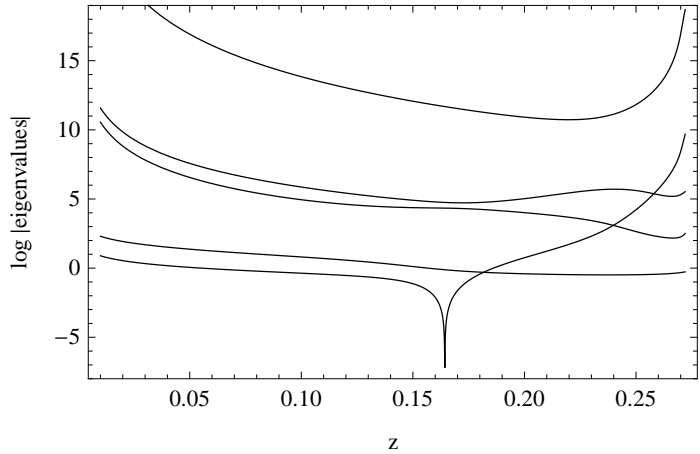}
\bigskip
\caption{The logarithm of the absolute value of the eigenvalues of the matrix of infinitesimal inequalities of order 5 as a function of $z$ (there are global factors depending on $R$ which we fix by setting $R_1=R_2=1$). One of the eigenvalues becomes negative for $z\gtrsim 0.165$.}
\label{auto1}
\end{figure}

\begin{figure}
\centering
\leavevmode
\epsfysize=6cm
\epsfbox{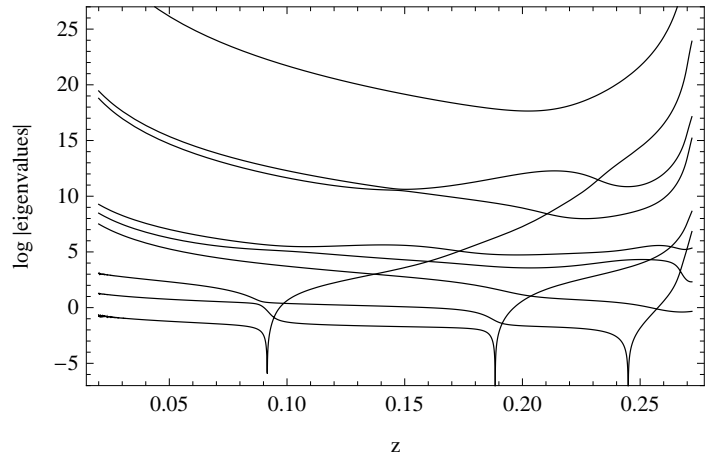}
\bigskip
\caption{The logarithm of the absolute value of the eigenvalues of the matrix of infinitesimal inequalities of order 9 as a function of $z$ (there are global factors depending on $R$ which we fix by setting $R_1=R_2=1$). There are negative eigenvalues for $z\gtrsim 0.09$.}
\label{auto2}
\end{figure}

 We are interested in the solution which describes a surface connecting the two loops. For each $z\le z_0=0.27288553...$ there are two solutions of 
equation (\ref{largui}) for $x$, but only one gives the minimal area. This corresponds to the solution with the greater value of $x$  (having $x\ge 0.58110028...$). The other 
solution corresponds to loops having the same orientation, which is not the case appearing in the reflection positivity inequalities (and it does not satisfy these inequalities).

Considering the case of circles of equal radius $R_1=R_2$ we have that the entropy is a one dimensional function and we evaluate the corresponding functions $f_n$. These are shown in figure (\ref{paraq}). This one parameter case seems to lead to a conditional positive function. 

Then, we consider the full two parameter case, and evaluate the eigenvalues of the matrices in the derivatives discussed in Section 4 with respect to both parameters, radius $R$ and distance $h$ to $x^0=0$ plane.  The first non trivial matrix (\ref{doe}) has 5 eigenvalues which we evaluate at the point of equal radius $R_1=R_2$. We plot these eigenvalues as a function of $z$ in figure (\ref{auto1}). We see this has a negative eigenvalue for $z\gtrsim 0.165$. Figure (\ref{auto2}) shows the eigenvalues of the matrices at the next order which consists of $9\times 9$ matrices. Negative eigenvalues appear now for $z\gtrsim 0.09$. According to the general theory, since the functions are definitely not conditionally positive in this case, the matrices of higher rank should be able to detect negative eigenvalues for $z$ as small as we want. 

The same analysis can be carried out for the case of two parallel concentric circular Wilson loops. The function of the cross ratio is the same, but this is given as $z=\frac{(h_1+h_2)^2+(R_1-R_2)^2}{4 R_1 R_2}\in (0,\infty)$ in terms of the radius $R_1$ and $R_2$ of the circles, and their distances $h_1$ and $h_2$ to the $x^0=0$ plane (which in this case is assumed parallel to the circles). The results are completely analogous to the ones above. In fact it can be seen that a conformal transformation commuting with the time reflection operation maps the positivity relations in one case and the other. Hence, we have infinite divisibility for the one parameter case of circles with equal radius, but not for the full two parameter function.

\subsection{Two spheres in four dimensions}

The case of two spheres can be treated similarly. Again the relevant part of the minimal area is conformally invariant and depends on a cross ratio $z = h \frac{(2 R_1 + 2 R_2 + h)}{4 R_1 R_2}\in (0,\infty)$. The formula for the area can be obtained from the result for two concentric shells \cite{hirata1} by a conformal transformation. We have for the mutual information of the two spheres $A$ and $B$,
\begin{eqnarray}
&&\log(1+2 z+2 \sqrt{z +z^2})=\frac{\cos^2(t_0)}{\sin^3(t_0)}\int_{0}^{\sin^2(t0)}dy\, \frac{y}{\sqrt{(1-y)((1-y)^2-y^3 \cos^4(t_0) /\sin^6(t_0)}}\,,\nonumber\\
&&I(A,B)=\lim_{\epsilon\rightarrow 0} \left[\frac{R_1^2+R_2^2}{2 \epsilon^2}-\frac{1}{2}\log\left(\frac{R_1 R_2}{\epsilon^2}\right)-\frac{1}{2}-\log(2)\right. \\
&&\left.-\int_{\frac{\epsilon}{R_1}}^{t_0} dt\, \frac{\cos^4(t)}{\sin^3(t)\sqrt{\cos^4(t)-\cos^4(t_0) \frac{\sin^6(t)}{\sin^6(t_0)}}}-\int_{\frac{\epsilon}{R_2}}^{t_0} dt\, \frac{\cos^4(t)}{\sin^3(t)\sqrt{\cos^4(t)-\cos^4(t_0) \frac{\sin^6(t)}{\sin^6(t_0)}}}\right]\,.\nonumber
\end{eqnarray} 
The connected minimal surface exists for $z\lesssim 0.0965$.

We obtain results very similar to the ones of the previous subsection. The one parameter function for equal radius $R_1=R_2$ gives place to (apparently) a conditional positive function. When the spheres are also allowed to vary in size the conditional positivity is lost, and we find negative eigenvalues for the matrices of multidimensional infinitesimal inequalities. We do not present the relevant figures here since they are in all respect similar to the ones of the last subsection for circular regions.  

\subsection{Multicomponent strips and the topology of the surfaces}
Now we consider regions formed by an arbitrary set of disjoint parallel strips. The analysis of infinite divisibility will display a curious phenomena, related 
to the Gross-Ooguri phase transition \cite{gross}. 

We take all long edges on the regions parallel to the $x^1$ direction, with size $L$. Let us call $a_1...a_n$ to the lines on the boundary with one orientation and $b_1...b_n$ to the lines with the opposite 
orientation, where $n$ is the number of connected components. In the large $L$ limit the relevant piece of the minimal surface will be some choice of all the 
possible surfaces formed by the union of the minimal surfaces connecting just two lines, one of type $a$ and one of type $b$. This can then be written as $(a_1,b_{\sigma(1)},a_2,b_{\sigma(2)},...,a_n,b_{\sigma(n)})$, with $\sigma(i)$, the permutation of $1...n$, producing the minimal area. This minimal area is just the sum ${\cal A}(a_1,b_{\sigma(1)})+...+{\cal A}(a_n,b_{\sigma(n)})$.     

Let us then think in a correlator involving $m$ of these multicomponent Wilson loops $V_{i}$, $i=1...m$. We are interested in the matrix ${\cal A}(V_{i}\bar{V}_j)$. Let $V_i$ have $2 n_i$ boundary lines. Then the minimal area of $V_i\bar{V}_{j}$ contains a group of $2q_{ij}\le 2 n_i$ surfaces which cross the $x^0=0$ plane and are connected to a group of $2 q_{ij}$ lines in $V_i$ and $2 q_{ij}$ lines in $\bar{V}_j$. There is also a group of $2 n_i-2 q_{ij}$ lines in $V_i$ which are connected among themselves, decoupled from $\bar{V}_{j}$. Then, suppose that in any of the correlators involving $V_i$ in the matrix, the same group of lines of $V_i$ are connected among themselves. The contribution of these lines is necessarily the same in all the correlators involving $V_i$ in the correlator matrix in question. Let us call this contribution $\xi_i$. Now suppose the same happens for all the regions $V_k$, that is, all the minimal surfaces in the correlator matrix elements $V_k \bar{V}_{j}$ which contain $V_k$ always has the same subset of the lines of $V_k$ which are connected among themselves. Then we write
\begin{equation}
{\cal A}(V_{i}\bar{V}_j)=\hat{{\cal A}}(V_{i}\bar{V}_j)+\xi_i+\xi_j\,,
\end{equation}   
where $\hat{{\cal A}}(V_{i}\bar{V}_j)$ contains the contribution from the surfaces which cross the $x^0=0$ plane. Eq. (\ref{tercio}) then tells that  $-{\cal A}(V_{i}\bar{V}_j)$ is conditionally positive if and only if   $-\hat{{\cal A}}(V_{i}\bar{V}_j)$ is conditionally positive. In this situation, the number of surfaces which cross $x^0$ must be the same in all correlator in the matrix, and hence $q_{ij}=q$ is independent of $i,j$. In this case the problem is simplified, we can forget about the disconnected pieces and think all lines are connected across $x^0=0$ by the minimal surface, and all loops have the same number of lines $2q$.   

Once we have disposed off the disconnected pieces, we prove infinite divisibility under the following conditions: For all $i,j$ we require there is a smooth transformation $V_{ij}(t)$, for $t\in [0,1]$, such that $V_{ij}(0)=V_i$, $V_{ij}(1)=V_j$. This requires in particular that all the regions have the same topology. Further, we ask these transformations to form a group under concatenation of mappings, $V_{ij}\circ V_{jk}=V_{ik}$. And finally, but most importantly,  we require that the sequence of minimal surfaces corresponding to $V_{ii^\prime}(t)\bar{V}_{jj^\prime}(t)$, going from the minimal surface of $V_{i}\bar{V}_{j}$ to the one of $V_{i^\prime}\bar{V}_{j^\prime}$, to be smooth.  That is, the trajectory of minimal surfaces  does not have to cross any phase transitions in the middle. In particular, the minimal surfaces on the correlator matrix must belong the same homotopy class of minimal surfaces.

It is immediate that this condition implies for the kind of loops we are considering here that the minimal surfaces connect only homologous lines. More clearly, let us name the lines in $V_i$ as $a^{(i)}_1,...,a^{(i)}_q$ and $b^{(i)}_1,...,b^{(i)}_q$. The conditions above imply we can choose the names of the lines such that in the minimal surface of $V_i\bar{V}_j$ the line $a^{(i)}_k$ is connected with $\bar{a}^{(j)}_k$, and $b^{(i)}_k$ with $\bar{b}^{(i)}_k$, for all $i,j,k$. With this naming of lines, all the minimal surfaces of the correlator matrix connect the $k^{th}$ lines of type $a$, and $b$ among themselves. Then we can write
\begin{equation}
-\bar{{\cal A}}(V_{i}\bar{V}_j)=-\sum_{k=1}^q ({\cal A}(a^{(i)}_k,\bar{a}^{(j)}_k)+{\cal A}(b^{(i)}_k,\bar{b}^{(j)}_k))\,.
\end{equation}  
This is a sum of conditionally positive matrices and then conditional positive. 

Conditional positivity cannot be proved in the case where the surfaces involved in the correlator matrix are separated by phase transitions, and some numerical examples show that generically it does not hold in these cases.   

Then, we can think that for more general cases than the one studied here the conditional positivity, except for the disconnected components, requires the minimal surfaces in the correlator matrix to lie in the same sector in the sense specified above. Notice that this does not exclude any topology for the minimal surfaces involved in the correlator. Generally, the conditions can always be achieved if the loops participating in the correlation matrix lie in some sufficiently small neighborhood of a given loop $V$. However, even in this cases without phase transitions we have already found some counterexamples of the inequalities in section 5.3. 

\section{Final comments}
Motivated by the positivity relations satisfied by the integer $n$ index Renyi entropies, and a natural conjecture about the path integral representation for the $n\rightarrow 1$ limit, we have checked whether the entanglement entropy gives place to conditionally positive correlation functions.  We could not find counterexamples of the inequalities (\ref{seiq}) among the few known exact QFT results. This is not so for the minimal areas of the holographic entanglement entropies where we found counterexamples for the case of two spheres or circles of different radius. In this case, the two parameter infinitesimal inequalities fail. The relation between analyticity and positivity then implies that the inequalities have to fail in the whole domain of holomorphy of the two variable correlator. We have also found phase transitions are a source of violations of the inequalities for multicomponent regions.  

A question which remains open is the status of the conditional positivity inequalities for the entropy in QFT. In the light of the results for minimal surfaces, the evidence in favor of the inequalities in QFT is quite weak. On one hand, most of the discussed cases refer to entanglement entropies for free theories, which have there own special properties and relations to the Renyi entropies. On the other, for the simple geometries analyzed, the minimal areas also obey the inequalities. The entanglement entropy for two spheres is not known exactly for a QFT yet. Of course, a positive answer for the infinite divisible property of the entanglement entropy in QFT would mean the holographic entropy ansatz has to be modified for some geometries. Perhaps the full solution of the entanglement entropy for a simple geometry such as the case of two intervals in general CFT would be enough to settle this question.
 
Another interesting question is the direct mechanism (either in the context of QFT or a purely geometric one) behind the validity of the inequalities in the one-parameter functions discussed in this paper. Indeed, we have found evidence indicating that several non trivial functions are conditional positive, obeying an infinite series of inequalities on the function derivatives. This includes minus the entropies of one interval and one angular sector for free scalar and fermions, and the minimal areas corresponding to angular sectors, and pairs of circles and spheres of equal radius. In terms of the minimal areas the conditional positivity property is rather peculiar. For example it is easy to check it does not hold for ordinary geodesics in AdS, but only for the limit of geodesics starting and ending at the boundary. 
 
One possible explanation could reside in the interplay between the positivity recurrence relations and some unknown differential recurrence relations for these functions. On the minimal surfaces these examples of infinite divisibility may be hinting to the validity for the small angle path integral representation of the entropy in these geometries. It would be particularly interesting to understand and generalize this enlargement of cases beyond the single sphere and the plane.   

It has been discussed in the literature that the small angle path integral representation is a weakness in Fursaev's proof of Ryu-Takayanagi proposal for holographic entanglement entropy \cite{holoc1,hedsolo}. We have further found it seems to involve  contradiction for certain geometries. On the QFT side the assumptions on which the proof is based lead to infinite divisibility, but the holographic result is in general not infinite divisible. However, for some geometries this contradiction is absent. 
 
\section*{Acknowledgments}
This work was supported by CONICET, Universidad Nacional de Cuyo, and CNEA, Argentina.
We were benefited by discussions with Juan Maldacena, Roger Melko, Rob Myers, Misha Smolkin, 
and Sergey Solodukhin. H.C. thanks the Aspen Center for Physics and the organizers of the Workshop ``Quantum information in quantum gravity and condensed matter physics'' for the kind invitation to participate and discuss some of the ideas in this paper.


\begin{thebibliography}{99}

\bibitem{wightman}
  R.~F.~Streater, A.~S.~Wightman,
  ``PCT, spin and statistics, and all that'',
  Redwood City, USA: Addison-Wesley (1989) (Advanced book classics).

\bibitem{melko} M. B. Hastings, I. Gonzalez, A. B. Kallin and R. G. Melko,  
``Measuring Renyi entanglement entropy with quantum monte carlo'', Phys. Rev. Lett. {\bf 104}, 157201 (2010) [arXiv:1001.2335]. 



\bibitem{replica}
  C.~G.~.~Callan and F.~Wilczek,
  ``On geometric entropy'', 
  Phys.\ Lett.\  B {\bf 333}, 55 (1994)
  [arXiv:hep-th/9401072].
  
\bibitem{replica1}
  F.~Larsen and F.~Wilczek,
  ``Geometric entropy, wave functionals, and fermions,''  Annals Phys.\  {\bf 243}, 280 (1995)  [hep-th/9408089].  

\bibitem{carcal}
  P.~Calabrese and J.~L.~Cardy,
  ``Entanglement entropy and quantum field theory,''  J.\ Stat.\ Mech.\ \ {\bf 0406}, P06002  (2004)  [hep-th/0405152].  

\bibitem{car}
  J.~L.~Cardy, O.~A.~Castro-Alvaredo and B.~Doyon,
   ``Form factors of branch-point twist fields in quantum integrable models and
  entanglement entropy'', 
  J.\ Statist.\ Phys.\  {\bf 130}, 129 (2008)
  [arXiv:0706.3384 [hep-th]].


\bibitem{rp} 
  K.~Osterwalder, R.~Schrader,
   ``Axioms For Euclidean Green's Functions'', 
  Commun.\ Math.\ Phys.\  {\bf 31}, 83-112 (1973).

\bibitem{you}
  H.~Casini,
   ``Entropy inequalities from reflection positivity'', 
  J.\ Stat.\ Mech.\  {\bf 1008}, P08019 (2010) 
  [arXiv:1004.4599 [quant-ph]].



  \bibitem{minkow}
  H.~Casini,
  ``Geometric entropy, area, and strong subadditivity,''
  Class.\ Quant.\ Grav.\  {\bf 21}, 2351 (2004)
  [arXiv:hep-th/0312238].

\bibitem{solre}
  S.~N.~Solodukhin,
  ``Entanglement entropy of black holes,''
  Living Rev.\ Rel.\  {\bf 14}, 8 (2011)
  [arXiv:1104.3712 [hep-th]].

\bibitem{holoc1}
  R.~C.~Myers, A.~Sinha,
   ``Holographic c-theorems in arbitrary dimensions'', 
  JHEP {\bf 1101}, 125 (2011) 
  [arXiv:1011.5819 [hep-th]].


\bibitem{solo}
  S.~N.~Solodukhin,
  ``Entanglement entropy, conformal invariance and extrinsic geometry'', 
  Phys.\ Lett.\  B {\bf 665}, 305 (2008)
  [arXiv:0802.3117 [hep-th]].


\bibitem{sphere}
  H.~Casini, M.~Huerta and R.~C.~Myers, 
  ``Towards a derivation of holographic entanglement entropy'', 
  JHEP {\bf 1105}, 036 (2011)
  [arXiv:1102.0440 [hep-th]].


\bibitem{sphere1}
  H.~Casini and M.~Huerta,
   ``Entanglement entropy for the n-sphere'', 
  Phys.\ Lett.\  B {\bf 694}, 167 (2010)
  [arXiv:1007.1813 [hep-th]]; 
  R.~Lohmayer, H.~Neuberger, A.~Schwimmer and S.~Theisen,
  ``Numerical determination of entanglement entropy for a sphere,''  Phys.\ Lett.\ B {\bf 685}, 222 (2010)  [arXiv:0911.4283 [hep-lat]];  
  J.~S.~Dowker,
  ``Entanglement entropy for even spheres,''  arXiv:1009.3854 [hep-th].  




\bibitem{myerscilin}
  L.~Y.~Hung, R.~C.~Myers and M.~Smolkin,
   ``On Holographic Entanglement Entropy and Higher Curvature Gravity'', 
  JHEP {\bf 1104}, 025 (2011)
  [arXiv:1101.5813 [hep-th]]; 
  J.~de Boer, M.~Kulaxizi and A.~Parnachev,
  ``Holographic Entanglement Entropy in Lovelock Gravities,''  JHEP {\bf 1107}, 109 (2011)  [arXiv:1101.5781 [hep-th]].  



\bibitem{cyl} 
  M.~Huerta,
  ``Numerical Determination of the Entanglement Entropy for Free Fields in the Cylinder,''  arXiv:1112.1277 [hep-th].  



\bibitem{fursaev}
  D.~V.~Fursaev,
   ``Proof of the holographic formula for entanglement entropy'', 
  JHEP {\bf 0609}, 018 (2006)
  [arXiv:hep-th/0606184].

\bibitem{ryu}
  S.~Ryu, T.~Takayanagi,
  ``Holographic derivation of entanglement entropy from AdS/CFT'', 
  Phys.\ Rev.\ Lett.\  {\bf 96}, 181602 (2006)
  [hep-th/0603001];
S.~Ryu and T.~Takayanagi,
 ``Aspects of holographic entanglement entropy'', 
JHEP {\bf 0608}, 045 (2006)
[arXiv:hep-th/0605073]. 


\bibitem{headrick}
  M.~Headrick, T.~Takayanagi,
   ``A Holographic proof of the strong subadditivity of entanglement entropy'', 
  Phys.\ Rev.\  {\bf D76}, 106013 (2007) 
  [arXiv:0704.3719 [hep-th]].

\bibitem{nishioka}
  T.~Nishioka, S.~Ryu and T.~Takayanagi,
   ``Holographic Entanglement Entropy: An Overview'', 
  J.\ Phys.\ A  {\bf 42}, 504008 (2009)
  [arXiv:0905.0932 [hep-th]].

\bibitem{hedsolo}
  M.~Headrick,
   ``Entanglement Renyi entropies in holographic theories'', 
  Phys.\ Rev.\  {\bf D82}, 126010 (2010) 
  [arXiv:1006.0047 [hep-th]].





\bibitem{hirata1}
  T.~Hirata, T.~Takayanagi,
   ``AdS/CFT and strong subadditivity of entanglement entropy'', 
  JHEP {\bf 0702}, 042 (2007) 
  [hep-th/0608213].

\bibitem{monogamy}
  P.~Hayden, M.~Headrick and A.~Maloney,
   ``Holographic Mutual Information is Monogamous'', 
  arXiv:1107.2940 [hep-th].




\bibitem{wilson}
  J.~M.~Maldacena,
   ``Wilson loops in large N field theories'', 
  Phys.\ Rev.\ Lett.\  {\bf 80}, 4859-4862 (1998) 
  [hep-th/9803002].

\bibitem{yaffe}
  L.~G.~Yaffe,
   ``Large n Limits as Classical Mechanics'', 
  Rev.\ Mod.\ Phys.\  {\bf 54}, 407 (1982).


\bibitem{matrices1} See for example R. Bhatia, ``Infinitely divisible matrices'', Am. Math. Monthly, {\bf 113}, 221 (2006).
 
\bibitem{matrices} See for example R. Bhatia, ``Positive definite matrices'', Princeton University Press, 
New Jersey (2007).

\bibitem{prob} See for example L. Breiman, ``Probability Theory'', Addison-Wesley (1968);
 B.V. Gnedenko, ``Theory of Probability'', 6th ed., Gordon and Breach (1997).
   
\bibitem{ot}
  M.~Moshe, J.~Zinn-Justin, 
  ``Quantum field theory in the large N limit: A Review'', 
  Phys.\ Rept.\  {\bf 385}, 69-228 (2003)
  [hep-th/0306133].

\bibitem{quark}
  C.~Bachas,
 ``Convexity Of The Quarkonium Potential'',
  Phys.\ Rev.\  {\bf D33}, 2723 (1986).


\bibitem{hirata2}
  T.~Hirata,
   ``New inequality for Wilson loops from AdS/CFT'', 
  JHEP {\bf 0803}, 018 (2008) 
  [arXiv:0801.2863 [hep-th]].


\bibitem{wedge}
  H.~Casini,
   ``Wedge reflection positivity'', 
  J.\ Phys.\ A {\bf A44}, 435202 (2011) 
  [arXiv:1009.3832 [hep-th]].

\bibitem{vander2}S. Izumi, ``Restrictions of smooth functions to a closed subset'', Ann. Inst. Fourier, Grenoble {\bf 54}, 1811 (2004) [math/0312226].

\bibitem{vander1}G. Galimberti and V. Pereyra, ``Numerical differenciation and the solution of multidimensional Vandermonde systems'', Mathematics of 
Computation {\bf 24}, 357 (1970).



\bibitem{review}  
  H.~Casini and M.~Huerta,
  ``Entanglement entropy in free quantum field theory,''  J.\ Phys.\ A A {\bf 42}, 504007 (2009)  [arXiv:0905.2562 [hep-th]].  

\bibitem{cari}
  O.~A.~Castro-Alvaredo and B.~Doyon,
  ``Bi-partite entanglement entropy in integrable models with backscattering,''  J.\ Phys.\ A A {\bf 41}, 275203 (2008)  [arXiv:0802.4231 [hep-th]];  
  O.~A.~Castro-Alvaredo and B.~Doyon,
  ``Bi-partite entanglement entropy in massive 1+1-dimensional quantum field theories,''  J.\ Phys.\ A A {\bf 42}, 504006 (2009)  [arXiv:0906.2946 [hep-th]].  



\bibitem{doyon}
  B.~Doyon,
  ``Bi-partite entanglement entropy in massive two-dimensional quantum field theory,''  Phys.\ Rev.\ Lett.\  {\bf 102}, 031602 (2009)  [arXiv:0803.1999 [hep-th]].  

\bibitem{data}
  H.~Casini, C.~D.~Fosco, M.~Huerta,
  ``Entanglement and alpha entropies for a massive Dirac field in two dimensions'', 
  J.\ Stat.\ Mech.\  {\bf 0507}, P07007 (2005) 
  [cond-mat/0505563].
  
  \bibitem{data1}
  H.~Casini, M.~Huerta,
  ``Entanglement and alpha entropies for a massive scalar field in two dimensions'', 
  J.\ Stat.\ Mech.\  {\bf 0512}, P12012 (2005) 
  [cond-mat/0511014]. 

\bibitem{hypo} 
  H.~Casini and M.~Huerta,
  ``Remarks on the entanglement entropy for disconnected regions,''  JHEP {\bf 0903}, 048 (2009)  [arXiv:0812.1773 [hep-th]].  


\bibitem{Calcato}
  P.~Calabrese, J.~Cardy and E.~Tonni,
  ``Entanglement entropy of two disjoint intervals in conformal field theory
  II,''
  J.\ Stat.\ Mech.\  {\bf 1101}, P01021 (2011)
  [arXiv:1011.5482 [hep-th]].
  P.~Calabrese, J.~Cardy and E.~Tonni,
  ``Entanglement entropy of two disjoint intervals in conformal field theory,''
  J.\ Stat.\ Mech.\  {\bf 0911}, P11001 (2009)
  [arXiv:0905.2069 [hep-th]].

\bibitem{Alba}
  V.~Alba, L.~Tagliacozzo and P.~Calabrese,
  ``Entanglement entropy of two disjoint intervals in c=1 theories,''
  J.\ Stat.\ Mech.\  {\bf 1106}, P06012 (2011)
  [arXiv:1103.3166 [cond-mat.stat-mech]];
  V.~Alba, L.~Tagliacozzo and P.~Calabrese,
  ``Entanglement entropy of two disjoint blocks in critical Ising models,'', Phys. Rev. {\bf B81}, 060411(R) (2010)  
  [arXiv:0910.0706 [cond-mat.stat-mech]].

\bibitem{Furukawa}
  S.~Furukawa, V.~Pasquier and J.~Shiraishi,
  ``Mutual Information and Compactification Radius in a c=1 Critical Phase in
  One Dimension,'', Phys. Rev. Lett. {\bf 102}, 170602 (2009)  
  [arXiv:0809.5113 [cond-mat.stat-mech]].

\bibitem{hea}
  M.~Headrick, A.~Lawrence and M.~M.~Roberts,
   ``Bose-Fermi duality and entanglement entropies,''  arXiv:1209.2428 [hep-th].  


\bibitem{gliozzi}
  M.~A.~Rajabpour and F.~Gliozzi,
  ``Entanglement Entropy of Two Disjoint Intervals from Fusion Algebra of Twist Fields,''  J.\ Stat.\ Mech.\  {\bf 1202}, P02016 (2012)  [arXiv:1112.1225 [hep-th]].  




\bibitem{Blanco}
  D.~D.~Blanco and H.~Casini,
  ``Entanglement entropy for non-coplanar regions in quantum field theory,''
  Class.\ Quant.\ Grav.\  {\bf 28}, 215015 (2011)
  [arXiv:1103.4400 [hep-th]].
  
  \bibitem{angul}
  H.~Casini, M.~Huerta,
  ``Universal terms for the entanglement entropy in 2+1 dimensions,''
  Nucl.\ Phys.\  {\bf B764}, 183-201 (2007) 
  [hep-th/0606256];
  H.~Casini, M.~Huerta, L.~Leitao,
  ``Entanglement entropy for a Dirac fermion in three dimensions: Vertex contribution,''
  Nucl.\ Phys.\  {\bf B814}, 594-609 (2009) 
  [arXiv:0811.1968 [hep-th]].

\bibitem{angulo}
  N.~Drukker, D.~J.~Gross, H.~Ooguri,
   ``Wilson loops and minimal surfaces'', 
  Phys.\ Rev.\  {\bf D60}, 125006 (1999) 
  [hep-th/9904191].

\bibitem{za}
  K.~Zarembo, 
  ``Wilson loop correlator in the AdS / CFT correspondence'', 
  Phys.\ Lett.\  {\bf B459}, 527-534 (1999) 
  [hep-th/9904149].


\bibitem{za1} 
  P.~Olesen, K.~Zarembo,
  ``Phase transition in Wilson loop correlator from AdS / CFT correspondence'', 
  hep-th/0009210.



\bibitem{tierri}
  N.~Drukker and B.~Fiol,
  ``On the integrability of Wilson loops in AdS(5) x S**5: Some periodic ansatze,''  JHEP {\bf 0601}, 056 (2006)  [hep-th/0506058].  


\bibitem{gross}
  D.~J.~Gross, H.~Ooguri,
  ``Aspects of large N gauge theory dynamics as seen by string theory'', 
  Phys.\ Rev.\  {\bf D58}, 106002 (1998) 
  [hep-th/9805129].


  
\end{thebibliography}
\end{document}